\documentclass[12pt]{iopart}
\usepackage{iopams}
\usepackage{epsf}
\usepackage{latexsym}
\usepackage[dvips]{graphicx}
\usepackage{epsfig}
\usepackage{amssymb}

\newcommand{\cut}[1]{{}}

\def\cN{{\cal N}}

\def\cA{{\cal A}}
\def\cL{{\cal L}}

\def\sN{{\mbox{\sf N}}}
\def\sD{{\mbox{\sf D}}}
\begin{document}
\title{Minimising Unsatisfaction in Colourful Neighbourhoods}

\author{
K. Y. Michael Wong{\ddag} and David Saad{\S}}
\address{\ddag\ Department of Physics,
Hong Kong University of Science \& Technology \\
Clear Water Bay, Hong Kong, China}
\address{\S\ Neural Computing Research Group, Aston University,
Birmingham B4 7ET, UK}

\begin{abstract}
Colouring sparse graphs under various restrictions is a theoretical
problem of significant practical relevance. Here we consider the
problem of maximising the number of different colours available at
the nodes and their neighbourhoods, given a predetermined number of
colours. In the analytical framework of a tree approximation,
carried out at both zero and finite temperatures, solutions
obtained by population dynamics
give rise to estimates
of the threshold connectivity
for the incomplete to complete transition,
which are consistent with those of existing algorithms.
The nature of the transition 
as well as the validity of the tree approximation 
are investigated.
\end{abstract}


\maketitle

\section{Introduction}

The spin glass theory of infinite-ranged
models~\cite{sherrington1975,kirkpatrick1978}
has inspired a generation of physicists
to study many theoretically challenging
and practically important problems
in physics and information processing~\cite{nishimori2001}.
These problems share a common feature,
in that the disordered interactions among their elements
cause frustration and non-ergodic behaviour.
The replica method~\cite{edwards1975} has been useful
in explaining their macroscopic behaviour.
At the same time,
based on the microscopic descriptions of the models,
the cavity method~\cite{mezard1987}
resulted in many computationally efficient schemes.
These approaches have laid the foundation
for the study of many problems in complex optimisation
using statistical mechanics, such as
graph partitioning~\cite{fu1986},
travelling salesman~\cite{mezard1986},
$K$-satisfiability~\cite{kirkpatrick1994},
and graph colouring~\cite{mulet2002}.

Not only the graph colouring problem~\cite{COLbook} 
is among the most basic NP-complete problems~\cite{NPbook},
but it also has direct
relevance to a variety of applications in scheduling, distributed
storage, content distribution and distributed computing.

In the original problem, one is given a graph and a number of
colours, and the task is to find a colouring solution such that
any two connected vertices are assigned different colours. This is
equivalent to the Potts glass with nearest neighbouring
interactions in statistical physics. The problem has been studied
by physicists using the cavity
method~\cite{mulet2002,braunstein2003}. For a given number of
colours, a phase transition takes place when the connectivity
increases, changing from a colourable to an uncolourable phase.
One of the statistical physics approaches was based on the replica
symmetric (RS) ansatz. It gave an over-estimate of the threshold
connectivity of this phase transition~\cite{vanmourik2002}. The
one-step replica symmetry-breaking (1RSB) approach takes into
account the possibility that the solution space can be
fragmented~\cite{mulet2002,braunstein2003}. Besides giving an
estimate of the threshold connectivity within the mathematical
bounds, it correctly predicts the existence of a clustering phase
below the threshold, in which the solution space spontaneously
divides into an exponential number of clusters. This is called the
hard colourable phase, in which local search algorithms are
rendered ineffective, and is a feature shared by other constraint
satisfaction problems~\cite{mezard2005,monasson1999}. The sequence
of phase transitions in the graph colouring problem, and their
algorithmic implications, were further refined
recently~\cite{krzakala2004,krzakala2006,zdeborova2007,zdeborova2007a}.

These advances in the spin glass theory
stimulated the development of efficient algorithms.
The cavity method gave rise to equations
identical to those of Belief Propagation (BP) algorithm
for graphical models~\cite{frey1998}.
Inspired by the 1RSB solution,
Survey Propagation (SP) algorithms
were subsequently developed to cope with situations
with fragmented solution space~\cite{mezard2002},
and they work well even in the hard phase
of the graph colouring problem~\cite{braunstein2003}.

In this paper, we study a variant
of the graph colouring problem, namely,
the colour diversity problem.
In this problem,
the aim is to maximise the number of colours
within one link distance of any node.
This is equivalent to the Potts glass
with second nearest neighbouring interactions
in statistical physics,
and hence is more complex than the original graph colouring problem
in terms of the increased number of frustrated links.
Indeed, this variant of the colouring problem
has been shown to be NP-complete~\cite{mccormick1983}.

This optimisation problem is directly related to various
application areas and in particular to the problem of distributed
data storage where files are divided to a number of segments,
which are then distributed over a graph representing the network.
Nodes requesting a particular file collect the required number of
file segments from neighbouring nodes to retrieve the original
information. Distributed storage is used in many real world
applications such as OceanStore~\cite{oceanstore}.

Compared with the original graph colouring problem,
work done on the colour diversity problem
mainly focused on algorithms~\cite{bounkong2006,jiang2007}.
Belief Propagation (BP) and Walksat algorithms for solving the
problem have been presented in~\cite{bounkong2006}.
Both algorithms revealed a transition
from incomplete to complete colouring,
and the possibility of a region of hard colouring
immediately below the transition point.
Approximate connectivity regimes for the solvable case
have been found, given the number of colours~\cite{bounkong2006}.
However, since the algorithms
are based on simplifying approximations (BP) and heuristics
(Walksat), both algorithms provide only upper bounds to the true
critical values.

The current study aims at providing a more principled approach to
study the problem, a theoretical estimate of the transition point,
and more insights on the nature of the transition itself. The method
employed is based on a tree approximation,
which is equivalent to the RS ansatz of the replica method
or the cavity method.
It results in a set of
recursive equations which can be solved analytically.
The connectivity values for
which the tree approximation is valid and the types of phases
present at each value are also investigated at both zero and
finite temperatures.

In section~\ref{sec:model} we introduce the model,
followed by section~\ref{sec:macro}
that explains briefly the derivation
and how the macroscopic behaviour can be studied.
In section~\ref{sec:results} we
present the results obtained via population dynamics.
Discussions on the behaviour at finite temperatures
are presented in section~\ref{sec:finitetemp} followed by a
concluding section.
The appendices contain further mathematical details.

\section{The Model}
\label{sec:model}

\subsection{The cost function}
Consider a sparsely connected graph with connectivity $c_i$ and
colour $q_i$ for node $i$.
The connectivities $c_i$ are drawn
from a distribution $P(c_i)$
with mean $\langle c\rangle$.
In this paper we consider the case of {\it linear connectivity},
that is, the nodes have connectivities
$\lfloor\langle c\rangle\rfloor$
or $\lfloor\langle c\rangle\rfloor+1$,
with probabilities $1-\langle c\rangle+\lfloor\langle c\rangle\rfloor$
and $\langle c\rangle-\lfloor\langle c\rangle\rfloor$ respectively.
The colour $q_i$ can take the values
$1,\cdots,Q$.
The colour diversity problem is trivial
for the case $\langle c\rangle>Q$,
in which colour schemes with complete sets of colours
available to all nodes can be found easily.
Hence we will focus on the more interesting case
$\langle c\rangle\le Q$,
in which a transition between
complete and incomplete colouring exists,
as shown in previous work~\cite{bounkong2006}.

The set of colours available at the node
and its local neighbourhood is
\[
    \cL_i  \equiv \{q_i\}\cup\{q_j|j\in N_i \},
\]
where $N_{i}$ is the set of nearest neighbours of node $i$.
To find a colour scheme that maximises
the number of different colours in $\cL_i$
and averaged over all nodes $i$,
we consider minimising the energy (cost function) of the form
\begin{equation}
\label{eq:cost}
    E=\sum\limits_i {\phi \left(\cL_i \right)}.
\end{equation}
Since the objective is equivalent to
minimising the number of identical colours in the set,
an appropriate form of the function $\phi$ is
\begin{equation}
    \phi\left(\cL_i\right)
    =\sum_{q_j\in\cL_i}\sum_{q_k\in\cL_i}
    \delta(q_j,q_k),
\end{equation}
where $\delta(a,b)=1$ for $a=b$, and 0 otherwise.
$\phi$ can be rewritten as
\begin{equation}
\label{eq:phi}
    \phi \left(\cL_i\right)
    =\sum\limits_{q=1}^Q {\left[ {\delta \left(
    {q,q_i } \right)+\sum\limits_{j\in N_i } {\delta \left( {q,q_j }
    \right)} } \right]} ^2.
\end{equation}

The quadratic nature of $\phi$ confirms that
it is an appropriate cost function
for diversifying the colours in the neighbourhood of each node.
Due to the convexity of its quadratic form,
its minimum solution tends to equalise
the numbers of all colours in the neighbourhood of a node.
Thus, besides maximising colour diversity,
our choice of the cost function
has an additional advantage for
the distributed storage optimisation task, which has motivated
the current study,
where an even distribution of segments (colours) in a neighbourhood
is also a secondary objective,
offering greater resilience.

The need for an even distribution of colours is especially important
when the total number of colours
is less than the connectivity of a node.
Consider the contribution from the function $\phi$
centred on a node in such a case.
Some colours can appear more than once.
Then the exact form of the function $\phi$
determines the selection of these extra colours.
In general, two types of selection can be made.
In the first type, one may still use all colours,
but they may be less evenly distributed than in the ground state.
In the second type, one may use fewer colours.
The former maximises the number of available colours,
but the latter does not.
In this case, an inappropriate choice of the cost function
will mix these two cases assigning the same energies,
rendering it impossible to distinguish
optimal and suboptimal colour choices.

On the other hand, Eq.~(\ref{eq:phi})
does not suffer from this shortcoming in the topology considered here.
A geometric interpretation is able to illustrate this point.
Let $n_q$ be the number of times colour $q$ appears in $\cL_i$.
Then the minimisation of Eq.~(\ref{eq:phi}) reduces to
the minimisation of $\sum_{q=1}^Q n_q^2$
subject to the constraint that $\sum_{q=1}^Q n_q=|N_i|+1$.
Note that the constraint defines a hyperplane
in the $Q$-dimensional space of $n_q$,
and the problem is equivalent
to finding the point with integer coordinates on the hyperplane
such that its distance from the origin is minimised.
The optimal solution is the point on the hyperplane
closest to the normal,
and no components should be zero when $Q\le|N_i|+1$.
In fact, the optimal solution is $n_q={\rm int}[(|N_i|+1)/Q]$
for $1\le q\le{\rm mod}(|N_i|+1,Q)$,
and $n_q={\rm int}[(|N_i|+1)/Q]+1$ otherwise
(or its permutations).

We have also considered a worst case analysis
of the change in the total cost
due to colour changes in neighbouring nodes when the function $\phi$, 
centred on a node $i$, is minimised.
It shows that for networks with linear connectivities
and $\langle c\rangle\le Q$,
the ground states consist of all satisfied nodes only,
if they exist.

\subsection{The statistical physics}
We note that second nearest neighbour interactions
are present in this cost function.
This is different from that of the original graph colouring problem,
where the cost function involves only nearest neighbour interactions.
As we shall see, the messages in the resultant message-passing algorithm
will be characterised by two components,
instead of the single components
in the case of the original graph colouring
problem~\cite{vanmourik2002,mulet2002}.

Analysis of the problem is done by writing
the free energy of the system at a temperature $T$, given by
\begin{equation}
    F=-T\ln Z,
\end{equation}
where $Z$ is the partition function given by
\begin{equation}
    Z={\rm Tr}_{\{q_i\}}\exp\left[-\beta\sum_i\phi(\cL_i)\right],
\end{equation}
$\beta\equiv T^{-1}$ being the inverse temperature.
In the zero temperature limit, the free energy
approaches the minimum cost function.
Several methods exist for deriving the free
energy based on the replica and tree-based approximations.
Here, the analysis adopts a tree-based approximation,
which is valid for sparse graphs.
When the connectivity of the graph is low,
the probability of finding a loop of finite length on the graph is low,
and the tree approximation well describes
the local environment of a node.
In the approximation, node $i$ is connected to $c_i$ branches
in a tree structure,
and the correlations among the branches of the tree are neglected.
In each branch, nodes are arranged in generations.
Node $i$ is connected to an ancestor node of the previous generation,
and another $c_i-1$ descendent nodes of the next generation.

\begin{figure}[htbp]
\centerline{\epsfig{height=7cm,file=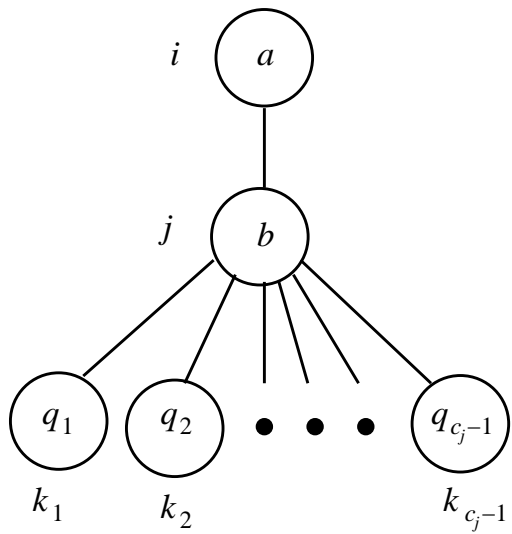}}
\caption{The notations used in computing
the vertex free energy $F^{V}_{ij}(a,b)$.
\label{fig:vertex}}
\end{figure}

Consider the free energy $F_{ij}(a,b)$ of the tree
terminated at node $j$ with colour $b$,
given its ancestor node $i$ of colour $a$.
In the tree approximation,
one notes that this free energy can be written as
$F_{ij}(a,b)=N_j F_{\rm av}+F_{ij}^V(a,b)$,
where $N_j$ is the number of nodes
in the tree terminated at node $j$,
and $F_{ij}^V(a,b)$ is referred to
as the {\it vertex free energy}~\cite{wong2006,wong2007}.
That is, the vertex free energy represents
the contribution of the free energy
extra to the average free energy
due to the presence of the vertex.
In the language of the cavity method,
$F_{ij}^V(a,b)$ are equivalent to the {\it cavity fields},
since they describe the state of the system
when node $i$ is absent.
The recursion relation of the vertex free energy of a node
can be obtained by considering the contributions
due to its descendent trees
and the energy centred at itself.
Using notations described in Fig.~\ref{fig:vertex},
the vertex free energy obeys the recursion relation
\begin{eqnarray}
\label{eq:vertexfreerecursion}
    F^{V}_{ij} (a,b)&=&-T\ln
    \mbox{Tr}_{\{q_k \vert k\in N_j \backslash \{i\}\}}
    \exp \Biggl[
    {-\beta \!\!\!\sum\limits_{k\in N_j \backslash \{i\}} {F_{jk}
    (b,q_k )} }  \\ &&- { \beta \phi \left( {b,\{a\}\cup \{q_k \vert
    k\in N_j \backslash \{i\}\}} \right)} \Biggr]-F_{\rm av} \ .
\nonumber
\end{eqnarray}
In the above expression,
the subtraction of $F_{\rm av}$
is due to the incorporation of node $j$
with the descendent trees
to form the tree terminated at node $j$.
For brevity, we will use the alternative simplified notation
\begin{equation}
\label{eq:vertexfreerecursion_al}
    F_{ij}^V (a,b)=-T\ln\mbox{Tr}_{\mathbf q}
    \exp \left[ -\beta \sum_{k=1}^{c_j -1} {F_{jk}^V
    (b,q_k )} -\beta \phi
    (a,b,{\mathbf q})\right]-F_{\rm av}  \ ,
\end{equation}
where the vector $\mathbf q$ refers to the colours of all descendants in
Fig.~\ref{fig:vertex}.

To find the average free energy $F_{\rm av}$,
one considers the contribution to a node $j$
due to all its $c_j$ neighbours, that is,
\begin{equation}
    F_{\rm av}=-T\left\langle\ln\mbox{Tr}_{\{\cL_i\}}
    \exp \left[ -\beta \sum\limits_{j\in N_i} {F_{ij}^V
    (b,q_k )} -\beta \phi(\cL_i)
    \right]\right\rangle_{\rm node}  \ ,
\label{eq:fav}
\end{equation}
where the average $\langle\cdots\rangle_{\rm node}$
denotes sampling of nodes with connectivity $c$
being drawn with probability $P(c)$.
However, since the probability of finding a descendant node
connecting to it is proportional to the number of links
the descendant has,
descendants are drawn with the {\it excess probability}
$cP(c)/\langle c\rangle$.

Equations (\ref{eq:vertexfreerecursion_al}) and (\ref{eq:fav})
can also be derived using the replica method
as presented in Appendix A.
We remark that both the derivation and the results
are very similar to those in the problem of resource allocation
on sparse networks~\cite{wong2006,wong2007},
where the dynamical variables are the real-valued currents
on the links of the networks.
The parallelism between resource allocation and colour diversity
is apparent when one notes that the currents in resource allocation
can be expressed as the differences between current potentials
defined on the nodes of the networks.
Hence the vertex free energies in both problems
can be considered as functions of two variables.

Another useful relation can be obtained
by substituting Eq.~(\ref{eq:vertexfreerecursion_al})
into Eq.~(\ref{eq:fav}),
\begin{equation}
    -T\left\langle\ln\mbox{Tr}_{a,b}\exp \left[
    -\beta F_{ij}^V(a,b)-\beta F_{ji}^V(b,a)
    \right]\right\rangle_{\rm link}=0  \ ,
\label{eq:link}
\end{equation}
where the average $\langle\cdots\rangle_{\rm link}$
denotes sampling of link vertices with connectivity $c$
with the excess probability.
This relation can be interpreted
by considering the free energy of forming a link
between vertices $i$ and $j$.
Since no extra nodes are added in this process,
the extra free energy should average to zero.

The average of a function $\cA(\cL_i)$ is given by
\begin{equation}
\label{eq:AvFree}
    \langle\cA\rangle
    =\left\langle\frac
    {\mbox{Tr}_{\{\cL_i\}}
    \exp\left[-\beta\sum\limits_{j\in N_i }{F^{V}_{ij}(q_i,q_j )}
    -\beta\phi\left(\cL_i\right)\right]\cA(\cL_i)}
    {\mbox{Tr}_{\{\cL_i\}}
    \exp\left[-\beta\sum\limits_{j\in N_i }{F^{V}_{ij}(q_i,q_j )}
    -\beta\phi\left(\cL_i\right)\right]}
    \right\rangle_{\rm node}.
\end{equation}
Hence the average energy is given by
\begin{equation}
    E_{\rm av} \equiv \langle E\rangle
    =\langle\phi\rangle_{\rm node}.
\label{eq:locale}
\end{equation}
The Edwards-Anderson order parameter $q_{\rm EA}$~\cite{gross1985},
whose nonzero value characterises the Potts glass phase,
is given by
\begin{equation}
    q_{\rm EA}=\frac{Q}{Q-1}\frac{1}{N}
    \sum_i\sum_q\left(\langle\delta(q,q_i)\rangle_{\rm node}
    -\frac{1}{Q}\right)^2.
\end{equation}
The performance measure of interest
is the {\it incomplete fraction} $f_{\rm incom}$,
which is defined as the average fraction of
nodes with an incomplete set of colours available at the node and
its nearest neighbours,
\begin{eqnarray}
\label{eq:incomplete}
    f_{\rm incom}
    =\left\langle\Theta \left[ {Q-\sum\limits_{q=1}^Q }{\Theta
    \left( {\delta (q,q_i )+\sum\limits_{j\in N_i } {\delta (q,q_j )}
    } \right)} \right] \right\rangle_{\rm node} ,
\end{eqnarray}
where $\Theta(q)=1$ for $q>0$, and 0 otherwise.
This performance measure is similar to the one used in \cite{bounkong2006},
which we refer to as the {\it unsatisfied fraction} $f_{\rm unsat}$,
and is defined as the average fraction of colours
unavailable at the node and its nearest neighbours
(for the case that $Q$ is not greater
than the number of nearest neighbours plus 1),
\begin{equation}
\label{eq:unsat}
    f_{\rm unsat}
    =\left\langle
    \left[ 1-\frac{1}{Q}\sum\limits_{q=1}^Q \Theta \left(
    {\delta (q,q_i )+\sum\limits_{j\in N_i } {\delta (q,q_j )} }
    \right) \right] \right\rangle_{\rm node}.
\end{equation}

One might consider using Eq.~(\ref{eq:incomplete}) or
(\ref{eq:unsat}) to define the cost function to be minimised,
instead of Eq.~(\ref{eq:phi}).
This is indeed possible
and we expect that zero-energy ground states can be obtained
when the condition of full colour diversity for each node is satisfiable.
In the unsatisfiable case, no zero-energy ground states can be found,
but one might still be interested
in finding states that minimise the average number of colours
unavailable to a node.
In this case, $f_{\rm incom}$ might not be an appropriate choice,
since it mixes up the energies
of selecting more (but unevenly distributed) colours,
and fewer colours.
The second measure
$f_{\rm unsat}$ favours those states
with higher colour diversity,
but for the same number of available colours,
it does not distinguish states
with different homogeneity of colour distribution.
By comparison, the cost function in Eq.~(\ref{eq:phi})
has the additional advantage of favouring homogeneous colour distributions
in the neighbourhood of the nodes.

\section{Macroscopic Properties}
\label{sec:macro}

\subsection{Population dynamics}
\label{subsec:popdyn}
Solutions to the recursive equation~(\ref{eq:vertexfreerecursion})
are obtained by population dynamics~\cite{mezard2001}.
We start with samples of $N$ nodes,
each with one of $Q$ colours
randomly assigned as the initial condition.
At each time step of the population dynamics, all the $N$ nodes
are updated once in random order.
At the instant we update node $j$,
we select $c_j-1$ nodes to be its descendants,
where $c_j$ is drawn from the distribution $P(c_{j})$.
Descendants with connectivities $c_k$ are randomly selected
with excess probabilities $c_kP(c_k)/\langle c\rangle$.
The vertex free energy is then
updated for all pairs $(a,b)$
{\it before} another node is updated.

We have also computed the solutions
using {\it layered} dynamics.
At each time step of the layered dynamics,
the new vertex free energies of all the $N$ nodes are calculated,
but are temporarily reserved
until the end of the time step.
Hence at the instant we renew node $j$,
we select $c_j-1$ nodes to be its descendants,
whose vertex free energies were computed in the {\it previous} time step.
Descendants with connectivities $c_k$ are randomly selected
with excess probabilities $c_kP(c_k)/\langle c\rangle$.
After the new vertex free energies of all the $N$ nodes have been computed,
they are then updated synchronously
and ready for the computation in the next time step.

We observe that a {\it modulation instability}
is present in layered dynamics~\cite{rivoire2004}.
This means that after sufficient layers of computation,
the colour distribution no longer remains uniform.
Rather, each layer is dominated by a particular colour,
and the dominant colour alternates from layer to layer.
This modulation is expected to be suppressed in random graphs
due to the presence of loops of incommensurate lengths.
Furthermore. the average free energy
computed by the layered dynamics
has variances increasing rapidly with layers.
Hence the layered dynamics is not adopted in our studies.

\subsection{Average free energy at finite temperatures}
To avoid growing fluctuations of the vertex free energies in the
population dynamics, their constant components are subtracted off
immediately after each update,
\mathindent = 3pc
\begin{equation}
\label{eq:Gdef}
    f_{ij}^V (a,b)\equiv F_{ij}^V (a,b)-G_{ij} \ ,
\end{equation}
where  $G_{ij}\equiv\sum_{c,d} {F_{ij}^V(c,d)}/Q^2$
is a constant bias independent of colours $a$ and $b$. The
recursion relation of the vertex free energy then becomes
\begin{equation}
\label{eq:iterations_f}
    f_{ij}^V (a,b)=-T\ln\mbox{Tr}_{\mathbf q}
    \exp \left[-\beta \sum\limits_{k=1}^{c_j -1} {f_{jk}^V
    (b,q_k)} -\beta \phi(a,b,{\mathbf q})
    \right]+\mbox{constant}
\end{equation}

After every time step, we measure the average free energy. This is
done by repeatedly creating a test node $j$ and randomly selecting
$c_j$ nodes to connect with the test node. The average free
energy is then given by
\begin{equation}
\label{eq:Fav1}
    F_{\rm av}=-\left\langle T\ln
    \mbox{Tr}_{\{\cL_i\}}
    \exp \left[ -\beta \sum_{j\in N_i}f_{ij}^V(q_i,q_j)
    -\beta \phi(\cL_i)\right]\right\rangle _{\rm node}
    \!\!\!\!\!\!+\langle c\rangle \langle G\rangle_{\rm link} \ .
\end{equation}
Note that $G$ is averaged over links,
since the descendants are drawn with excess probabilities.
To calculate $\langle G\rangle _{\rm link}$ we employ the
consistency condition (\ref{eq:link})
for the average free energy of a link,
which requires
\begin{equation}
\label{eq:link_normalization}
    -\left\langle {T\ln\mbox{Tr}_{a,b}
    \exp \left[ {-\beta f_{ij}^V (a,b)-\beta f_{ji}^V
    (b,a)} \right]}
    \right\rangle _{\rm link} \!\! +2\left\langle G
    \right\rangle _{\rm link} = 0.
\end{equation}
The node and link samplings are identical
for graphs with uniform connectivity.
This allows us to eliminate $\langle G\rangle$
in Eqs.~(\ref{eq:Fav1}) and (\ref{eq:link_normalization}),
and thus obtain $F_{\rm av}$.
To tackle the case of non-uniform connectivities,
we need to generalise the consistency condition
(\ref{eq:link_normalization}).
This can be done by restricting
our consideration to links with vertices of given
connectivities $A$ and $B$, and consider the free energy
due to the link connecting the trees on both sides of such links
\mathindent = 3pc
\begin{equation}
\label{eq:link_normalization2}
    -\left\langle {T\ln \mbox{Tr}_{a,b}
    \exp \left[ {-\beta F_{ij}^V (a,b)-\beta F_{ji}^V (b,a)}
    \right]}\right\rangle _{C_i =A,C_j =B} =0 \ .
\end{equation}
The derivation is analogous to that of
Eq.~(\ref{eq:link_normalization}), resulting in
\begin{equation}
\label{eq:GaGb}
    -\left\langle {T\ln \mbox{Tr}_{a,b} \exp \left[
    {-\beta f_{ij}^V (a,b)-\beta f_{ji}^V (b,a)} \right]}
    \right\rangle _{C_i =A,C_j =B} +\left\langle G
    \right\rangle _A
    +\left\langle G \right\rangle _B =0,
\end{equation}
which facilitates the elimination of the biases $G$
in Eq.~(\ref{eq:Fav1}),
resulting in an expression for the average free energy
\mathindent = 1 pc
\begin{eqnarray}
\label{eq:Fav}
    F_{\rm av}
    &=& -\left\langle T\ln
    \mbox{Tr}_{\{\cL_i\}}
    \exp \left[ -\beta \sum_{j\in N_i}f_{ij}^V(q_i,q_j)
    -\beta \phi(\cL_i)\right]\right\rangle _{\rm node}
    \\
    &&+\frac{\langle c\rangle}{2}\sum_{A,B}
    \frac{AP(A)}{\langle c\rangle}\frac{BP(B)}{\langle c\rangle}
    \left\langle T\ln \mbox{Tr}_{a,b}
    \exp \left[ -\beta f_{ij}^V (a,b)-\beta f_{ji}^V (b,a) \right]
    \right\rangle _{C_i=A,C_j=B}.\nonumber
\end{eqnarray}
\mathindent = 6pc

To evaluate  $F_{\rm av}$ one first performs the node average in
the first term of Eq.~(\ref{eq:Fav}), keeping a record of the
number of times each node $k$ is sampled. Then one performs the
average in the second term, randomly drawing the vertices $i$ and
$j$ of the links from nodes $k$ with exactly the same number of
times they appear in the first term.
Hence in this procedure,
the descendants in both terms are drawn from the excess distribution.
Furthermore, it ensures that the $G_{ij}$'s
appearing in the first term are exactly cancelled by those
appearing in the second term, thus eliminating a source of
possible fluctuations.

We also note that there can be a variety of choices of $G_{ij}$'s
to be subtracted from the vertex free energies
in Eq.~(\ref{eq:Gdef}).
For example, one may choose $G_{ij}$ to be $F_{ij}^V(1,1)$
and arrive at the same result Eq.~(\ref{eq:Fav}).
In fact, this computationally simple choice
is adopted in our computation.

\subsection{Energy and entropy at finite temperatures}
\label{subsec:esfinite}
Expressions for the energy and entropy follow immediately using
the identity $E\!=\! \partial{(\beta F})/\partial \beta $ and the
averaging of Eq.~(\ref{eq:AvFree}),
\mathindent = 1pc
\begin{eqnarray}
\label{eq:Eav}
    E_{\rm av} &=&\left\langle\frac
    {{\rm Tr}_{\{\cL_i\}}
    \exp \left[-\beta \sum_{j\in N_i} F_{ij}^V (q_i,q_j)
    -\beta\phi(\cL_i)\right]
    \left[\sum_{j\in N_i} E_{ij}^V (q_i,q_j)
    +\phi(\cL_i)\right]}
    {{\rm Tr}_{\{\cL_i\}}
    \exp \left[-\beta \sum_{j\in N_i} F_{ij}^V (q_i,q_j)
    -\beta\phi(\cL_i)\right]}
    \right\rangle_{\rm node},
    \nonumber\\
\end{eqnarray}
where $E_{ij}^V (a,b)$ is the vertex energy with the recursion
relation
\begin{eqnarray}
\label{eq:Eav2}
    E_{ij}^V(a,b)&=&\frac{\mbox{Tr}_{\mathbf q}
    \exp \left[-\beta \sum_{k=1}^{c_j-1} F_{jk}^V(b,q_k)
    -\beta\phi(a,b,{\mathbf q})\right]
    \left[\sum_{k=1}^{c_j-1} E_{jk}^V (b,q_k)
    +\phi(a,b,{\mathbf q})\right]}
    {\mbox{Tr}_{\mathbf q}
    \exp \left[-\beta \sum_{k=1}^{c_j-1} F_{jk}^V(b,q_k)
    -\beta\phi(a,b,{\mathbf q})\right]}
    \nonumber\\
    &&-E_{\rm av} \ ,
\end{eqnarray}
and
\mathindent = 6pc
\begin{equation}
\label{eq:entropy}
    S=\frac{E_{\rm av} -F_{\rm av} }{T} \ .
\end{equation}

Compared with the previous equation (\ref{eq:locale})
for the average energy,
Eq.~(\ref{eq:Eav}) includes the vertex energies of the descendants.
These vertex energies transmit the energy deviations
from the average energy, from the descendants to the ancestors.
Hence Eq.~(\ref{eq:Eav}) can be regarded as a global estimate
of the average energy,
and Eq.~(\ref{eq:locale}) is a local estimate.
Theoretically, one expects that both estimates
should yield the same result.
Numerically, however, we found that this is only valid
in the paramagnetic phase.
In the Potts glass phase,
the discrepancy between the two estimates
can be very significant.
This shows that in the paramagnetic phase,
memories about the initial conditions are lost easily.
In contrast, in the Potts glass phase,
memories about the initial conditions can propagate
for a long time through the vertex energies.

To avoid propagating fluctuations
in the computation of the average energy,
we subtract $E_{ij}^V(1,1)$ from all components $E_{ij}^V(a,b)$
immediately after each update,
and find $E_{\rm av}$ using
\mathindent = 1 pc
\begin{eqnarray}
    E_{\rm av}&=&\left\langle\frac
    {\mbox{Tr}_{\{\cL_i\}}
    \exp \left[-\beta \sum_{j\in N_i} f_{ij}^V (q_i,q_j)
    -\beta\phi(\cL_i)\right]
    \left[\sum_{j\in N_i} E_{ij}^V (q_i,q_j)
    +\phi(\cL_i)\right]}
    {\mbox{Tr}_{\{\cL_i\}}
    \exp \left[-\beta \sum_{j\in N_i} f_{ij}^V (q_i,q_j)
    -\beta\phi(\cL_i)\right]}
    \right\rangle_{\rm node}
    \nonumber\\
    &&-\frac{\langle c\rangle}{2}\sum_{A,B}
    \frac{AP(A)}{\langle c\rangle}\frac{BP(B)}{\langle c\rangle}
    \nonumber\\
    &&\times\left\langle\frac
    {\mbox{Tr}_{a,b}
    \exp \left[-\beta f_{ij}^V(a,b)-\beta f_{ji}^V(b,a)\right]
    \left[ E_{ij}^V(a,b)+E_{ji}^V(b,a)\right]}
    {\mbox{Tr}_{a,b}
    \exp \left[-\beta f_{ij}^V(a,b)-\beta f_{ji}^V(b,a)\right]}
    \right\rangle_{c_i=A,c_j=B}.
\label{eq:globale}
\end{eqnarray}

\subsection{Free energy, energy and entropy at zero temperature}
\label{zerotemp}
The derivation at zero temperature should be carried
out with extra care due to possible degeneracy in the
solutions. In the zero temperature
limit, Eq.~(\ref{eq:vertexfreerecursion_al}) reduces to
\mathindent = 6 pc
\begin{equation}
\label{eq:vertexfreerecursion_al0}
    F_{ij}^V (a,b)=\min_{\mathbf q}
    \left[\sum_{k=1}^{c_j -1} {F_{jk}^V (b,q_k )}
    +\phi(a,b,{\mathbf q})\right]
    -F_{\rm av}  \ .
\end{equation}
The expression of the entropy at zero temperature can be computed
directly from the \emph{vertex entropies}.
Differentiating Eq.~(\ref{eq:vertexfreerecursion_al})
with respect to $T$, and taking the zero
temperature limit, one obtains
\begin{equation}
    S_{ij}^V (a,b)=\ln \left[
    \sum_{\{{\mathbf q}^*\}}\exp \left(\sum_{k=1}^{c_j-1}S_{jk}^V
    (b,q_k^* )\right)\right]-S_{\rm av},
\end{equation}
where $\{{\mathbf q}^*\}$
is the set of colours minimising the free energy
$\sum_{k=1}^{c_j}F_{jk}^V (b,q_k) +\phi(a,b,{\mathbf q})$
at node $j$.
Similarly, differentiating Eq.~(\ref{eq:Fav}) with respect to $T$
and taking the zero temperature limit, one obtains
\mathindent = 1 pc
\begin{eqnarray}
    S_{\rm av} &=& \left\langle\ln\left[\sum_{\{\cL_i^*\}}
    \exp\left(\sum_{j\in N_i} S_{ij}^V(q_i^*,q_j^*)\right)\right]
    \right\rangle _{\rm node}
    \\ &-&\frac{\langle c\rangle}{2}\sum_{A,B}
    \frac{AP(A)}{\langle c\rangle}\frac{BP(B)}{\langle c\rangle}
    \left\langle {\ln \left[ {\sum\limits_{\{a^* ,b^* \}}
    {\exp\left( {S_{ij}^V (a^* ,b^* )+S_{ji}^V (b^* ,a^*)} \right)} }
    \right]} \right\rangle _{c_i=A,c_j=B} \nonumber \ ,
\end{eqnarray}
where $\{\cL_i^*\}$ are the set of colours minimising the free energy
$\sum_{j\in N_i} F_{ij}^V (q_i,q_j) +\phi(\cL_i)$
at node $i$, and $\{a^*, b^*\}$ are the set of the pair of colours
minimising the free energy $F_{ij}^V (a,b)+F_{ji}^V (b,a)$ at link
\textit{ij}.

The performance measures are now weighted by the entropies, and
Eq.~(\ref{eq:AvFree}) is replaced by the expression
\mathindent = 6 pc
\begin{equation}
\label{eq:AvEnt}
    \langle\cA\rangle
    =\left\langle\frac
    {\mbox{Tr}_{\{\cL_i^*\}}
    \exp\left[\sum\limits_{j\in N_i }{S_{ij}(q_i^*,q_j^*)}\right]\cA(\cL_i^*)}
    {\mbox{Tr}_{\{\cL_i^*\}}
    \exp\left[\sum\limits_{j\in N_i }{S_{ij}(q_i^*,q_j^*)}\right]}
    \right\rangle_{\rm node}.
\end{equation}

\subsection{The paramagnetic state at finite temperatures}
\label{subsec:para}
In the paramagnetic state,
the vertex free energies are symmetric
with respect to permutation of colours at each node.
Hence there are only two distinct values
of the vertex free energy for each node,
corresponding to the cases that the colours of the node and its ancestor
are the same or different.
Hence, we can derive the recursion relation for the single variable
$z_{ij}\equiv\exp[-\beta(F_{ij}^V(a,a)-F_{ij}^V(a,b))]$,
where $a\ne b$.
This is a significant simplification
of the original recursion relation for $F_{ij}^V(a,b)$,
which involves $Q^2$ components.

Specifically, we consider graphs with linear connectivity
$3\le\langle c\rangle\le 4$.
We first consider the vertex free energy
of a node $j$ with $c_j=3$,
whose descendants are labelled 1 and 2.
The recursion relations are given by
\mathindent = 1 pc
\begin{eqnarray}
\label{eq:para_rec}
    F_{ij}^V(a,a) &=& -T\ln \mbox{Tr}_{q_1 ,q_2 } \exp \left[
    {-\beta F_{j1}^V (a,q_1 )-\beta F_{j2}^V (a,q_2 )-\beta \phi
    (a,a,q_1 ,q_2 )} \right]-F_{\rm av},
    \nonumber\\
    F_{ij}^V(a,b) &=& -T\ln \mbox{Tr}_{q_1 ,q_2 } \exp \left[
    {-\beta F_{j1}^V (a,q_1 )-\beta F_{j2}^V (b,q_2 )-\beta \phi
    (a,b,q_1 ,q_2 )} \right]-F_{\rm av}.
\end{eqnarray}
By explicitly tabulating the different colour configurations
and introducing the notations $z\equiv\exp(-\beta)$
and $Q_n\equiv Q-n$, one can rewrite Eq.~(\ref{eq:para_rec}) as
\begin{eqnarray}
\label{eq:para_linear_zij}
    F_{ij}^V (a,a) &=& -T\ln \left[
    z^{16}z_{j1} z_{j2}+Q_1z^{10}(z_{j1} +z_{j2})+Q_1z^8+Q_1Q_2z^6 \right]
    \nonumber \\
    &&+\sum_k F_{jk}^V(a,b)-F_{\rm av},  \nonumber \\
    F_{ij}^V (a,b)&=& -T\ln \left[
    z^{10}z_{j1} z_{j2}+(z^8+Q_2z^6)(z_{j1} +z_{j2})
    +z^{10}+3Q_2z^6+Q_2Q_3z^4\right]
    \nonumber\\
    &&+\sum_k F_{jk}^V(a,b)-F_{\rm av}.
\end{eqnarray}
These give rise to the recursion relation for $z_{ij}$,
\begin{equation}
    z_{ij} = z^2 \left(
    {\frac{Q_1Q_2+Q_1z^2+Q_1z^4(z_{j1} +z_{j2} )+z^{10}z_{j1}z_{j2} }
    {Q_2Q_3+3Q_2z^2+z^6+(Q_2z^2+z^4)(z_{j1} +z_{j2})+z^6z_{j1} z_{j2}}}
    \right).
\label{eq:recursion_z}
\end{equation}
Similarly, for node $j$ with $c_j=4$,
\begin{equation}
    z_{ij} =z^2\left(\frac{Z_{\rm N}}{Z_{\rm D}}\right),
\label{eq:recursion_z4}
\end{equation}
where
\begin{eqnarray}
\label{eq:para_z4}
    Z_{\rm N}
    &=& Q_1Q_2Q_3+3Q_1Q_2z^2+Q_1z^6
    + (Q_1Q_2z^4+Q_1z^6)(z_{j1} +z_{j2} +z_{j3} )
    \nonumber\\
    &+& Q_1z^{10}(z_{j1} z_{j2} +z_{j2} z_{j3} +z_{j1} z_{j3})
    +z^{18}z_{j1} z_{j2} z_{j3} \ ,
    \nonumber \\
     Z_{\rm D}&=& Q_2Q_3Q_4+6Q_2Q_3z^2+3Q_2z^4+4Q_2z^6+z^{12}
    \nonumber \\
    &+& (Q_2Q_3z^2+3Q_2z^4+z^8)(z_{j1} +z_{j2}+z_{j3})
    \nonumber \\
    &+& (Q_2z^6+z^8)(z_{j1} z_{j2} +z_{j2} z_{j3}+z_{j1} z_{j3})
    +z^{12}z_{j1} z_{j2} z_{j3} \ .
\end{eqnarray}
Expressions of the average free energy and average energy
can be found in Appendix B.

\subsection{The paramagnetic state at zero temperature}
In the zero temperature limit for $Q\le 4$,
Eqs.~(\ref{eq:recursion_z}) and (\ref{eq:recursion_z4}) reduce to
\begin{eqnarray}
    &&z_{ij}=\left(\frac{Q_1}{Q_3}\right)z^2\to 0\quad{\rm for\ }c_j=3,
    \nonumber\\
    &&z_{ij}=\frac{Q_1}{6+z_{j1}+z_{j2}+z_{j3}}
    \quad{\rm for\ }c_j=4.
\end{eqnarray}
For $c_j=4$, the range of values of $z_{ij}$
is $2Q_1/(Q_1+12)\le z_{ij}\le Q_1/6$.
Hence the distribution of the vertex partition function
is given for $c_j=4$ by
\begin{equation}
    P(z)=\sum_{k=0}^3
    \left(\begin{array}{c}3\\k\end{array}\right)
    f_3^{3-k}f_4^k
    \prod_{r=1}^k\left[\int_{2Q_1/(Q_1+12)}^{Q_1/6}
    dz_r P(z_r)\right]
    \delta\left(z-\frac{Q_1}{6+\sum_{r=1}^k z_r}\right),
\label{eq:pz}
\end{equation}
where
$f_3\equiv 3(4-\langle c\rangle)/\langle c\rangle$
and $f_4\equiv 4(\langle c\rangle-3)/\langle c\rangle$
are the excess probabilities,
and are distinctive from the connectivity probabilities
$p_3\equiv 4-\langle c\rangle$ and $p_4\equiv\langle c\rangle-3$
in subsequent expressions.

For the average free energy, Eq.~(\ref{eq:para4Fav}) becomes
\begin{eqnarray}
    \left. F_{\rm av}\right|_{c=3}
    &=& 4-T\ln(Q Q_1 Q_2 Q_3),
    \nonumber \\
    \left. F_{\rm av}\right|_{c=4}
    &=& 7-T\sum_{k=0}^4
    \left(\begin{array}{c}4\\k\end{array}\right)
    f_3^{4-k}f_4^k
    \prod_{r=1}^k\left[\int_{2Q_1/(Q_1+12)}^{Q_1/6}
    dz_r P(z_r)\right]
    \nonumber\\
    &&\times\ln\left[Q Q_1 Q_2 Q_3
    \left(6+\sum_{r=1}^k z_r\right)\right],
    \nonumber\\
    \left.F_{\rm link}\right|_{C_1C_2}
    &=&-(1-\delta_{C_1,4}\delta_{C_2,4})T(1-f_4^2)\ln Q Q_1
    -\delta_{C_1,4}\delta_{C_2,4}
    Tf_4^2\int_{2Q_1/(Q_1+12)}^{Q_1/6}dz_1 P(z_1)
    \nonumber\\
    &&\times\int_{2Q_1/(Q_1+12)}^{Q_1/6}dz_2 P(z_2)
    \ln\left[Q(Q_1+z_1z_2)\right].
\end{eqnarray}
Hence in the zero temperature limit,
\begin{equation}
    F_{\rm av}=3\langle c\rangle-5.
\end{equation}
This value of the average free energy interpolates between 4 and 7
at $\langle c\rangle = 3$ and 4 respectively.
This means that in the paramagnetic phase,
there is a freedom in assigning the colours of the nodes
so that all local energies are minimised.
For a node with 3 neighbours and $Q=4$,
the state of local energy minimum
has one of each colour among itself and its neighbours.
Hence the energy is 4.
Similarly, for a node with 4 neighbours and $Q=4$,
the state of local energy minimum has,
among itself and its neighbours,
two nodes of the same colour
and three nodes of mutually different colours.
Hence the energy is 7.
The result of $3\langle c\rangle-5$ is the average of 4 and 7,
weighted by the fraction of nodes
with 3 and 4 neighbours respectively.
This is the lowest possible energy of the system.

The average entropy of the paramagnetic state is given by
\begin{eqnarray}
    S_{\rm av}
    &=& \ln(Q Q_1 Q_2 Q_3)+p_4f_3\ln Q_1
    -\frac{\langle c\rangle}{2}[\ln(Q Q_1)-f_4^2\ln Q_1]
    \nonumber\\
    &&-p_4\int_{2Q_1/(Q_1+12)}^{Q_1/6}dz P(z)\ln z
    -\left(\frac{\langle c\rangle}{2}f_4-p_4\right)f_4
    \int_{2Q_1/(Q_1+12)}^{Q_1/6}dz_1 P(z_1)
    \nonumber\\
    &&\times\int_{2Q_1/(Q_1+12)}^{Q_1/6}dz_2 P(z_2)
    \ln[Q(Q_1+z_1z_2)].
\label{eq:sav}
\end{eqnarray}
Consider the case $Q=4$.
When $\langle c\rangle = 3$, $S_{\rm av} = -\ln 3/2$.
For general values of $Q$,
we have $S_{\rm av} = \ln(Q_2 Q_3/\sqrt{Q Q_1})$.
Hence the entropy becomes negative for $Q = 4$,
although the entropy remains positive for $Q > 4$.

On the other hand, when $\langle c\rangle=4$,
the vertex partition function becomes node independent,
implying $z=\sqrt 2 -1$,
and $S_{\rm av}=\ln[(15+12\sqrt 2)/28]=0.13$.
Hence at an intermediate value of $\langle c\rangle$,
the entropy changes sign.
Thus there is a range of negative entropy
for $\langle c\rangle$ below 4
where the RS ansatz is unstable.

\section{Numerical results}
\label{sec:results}

Numerical solutions to the equations are obtained using population
dynamics in the manner explained in
subsection~\ref{subsec:popdyn}.
Results are obtained for $Q=4$
and ensembles of graphs with linear connectivity
$3\le\langle c\rangle\le 4$,
mixing nodes with connectivities 3 and 4 in varying proportions.
After every time step, we measure the following measures:
the local estimate of the average energy, the incomplete fraction,
and the Edwards-Anderson order parameter.
This is done by creating a test node $i$ and
randomly selecting $c_i$ nodes to connect with the test node.
The node contributions to the average free energy,
the global estimate of the average energy,
and (for zero temperature) the entropy
are also computed.
The computed measures are repeated for $N = 10000$ nodes
for each sample.
The set of descendant nodes of these $N$ test nodes is recorded.
Then, pairs of nodes are randomly drawn this set to form links,
and the link contributions to the average free energy,
the global estimate of the average energy,
and (for zero temperature) the entropy are computed.

\subsection{Paramagnetic and Potts glass phases}
\label{subsec:para-glass}
\begin{figure}[ht]
\centerline{\epsfig{height=7cm,figure=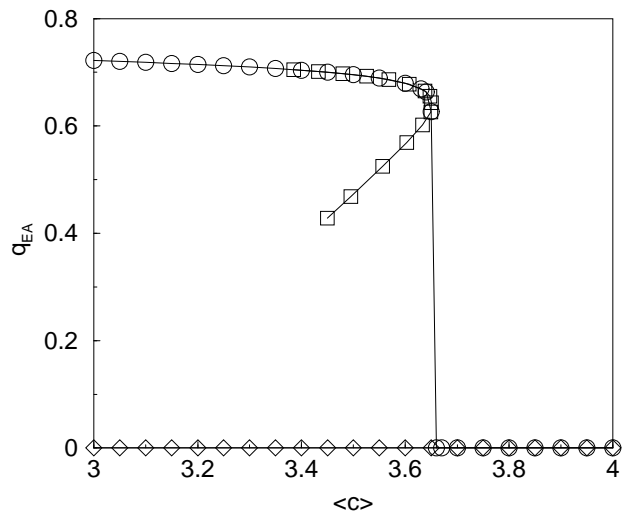}}
\caption{
The dependence of the Edwards-Anderson order parameter $q_{\rm EA}$
on the average connectivity $\langle c\rangle$,
obtained from the population dynamics
at fixed $\langle c\rangle$ ($\bigcirc$),
at fixed $f_{\rm incom}$ ($\square$)
and for the paramagnetic state ($\lozenge$).
Parameters: $N=10000$, $Q=4$ and 30 samples.
\label{fig:qea}}
\end{figure}

Figure~\ref{fig:qea} shows the Edwards-Anderson order parameter
as a function of $\langle c\rangle$.
It can be seen that the value of $q_{\rm EA}$ is 0
in the paramagnetic phase, which spans the region
$\langle c\rangle\ge\langle c\rangle_{\rm sp}=3.65$.
In this phase, all nodes have free choices of colours.
The Potts glass phase spans the region
$\langle c\rangle<\langle c\rangle_{\rm sp}$,
where $q_{\rm EA}$ remains at a value around 0.7,
and its transition to the paramagnetic phase
is of the first order.

\begin{figure}[ht]
\centerline{\epsfig{height=7cm,figure=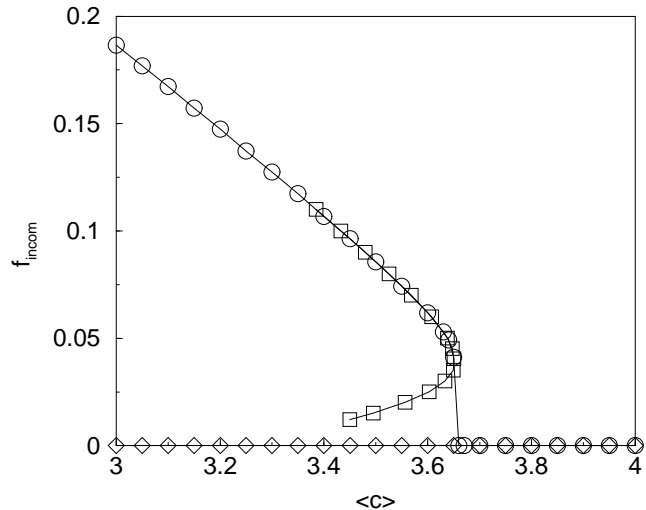}}
\caption{
The dependence of the incomplete fraction $f_{\rm incom}$
on the average connectivity $\langle c\rangle$.
Symbols and parameters: as in Fig.~\ref{fig:qea}.
\label{fig:fincom}}
\end{figure}

Figure~\ref{fig:fincom} shows incomplete fraction obtained
from the steady state solution of the population dynamics at fixed
$\langle c\rangle $ values.
It remains nonzero in the Potts glass phase,
and vanishes discontinuously above $\langle c\rangle_{\rm sp}$
in the paramagnetic phase.
To find the stable as well as the unstable solutions
of the population dynamics,
which correspond to multiple solutions
at fixed $\langle c\rangle$,
we may run the population dynamics
at fixed nonzero $f_{\rm incom}$.
This can be done by monitoring $f_{\rm incom}$
conditionally averaged on the nodes with
$c_j=\lfloor\langle c\rangle\rfloor$ and
$c_j=\lfloor\langle c\rangle\rfloor+1$ at each step,
and adjusting the value of $\langle c\rangle$ to approach
its targeted value, which is related to
the targeted value of $f_{\rm incom}$
estimated at each time step by
$f_{\rm incom}=(\langle c\rangle-\lfloor\langle c\rangle\rfloor)
f_{\rm incom}|_{c=\lfloor\langle c\rangle\rfloor+1}
+(1-\langle c\rangle+\lfloor\langle c\rangle\rfloor)
f_{\rm incom}|_{c=\lfloor\langle c\rangle\rfloor}$.
The population dynamics at fixed $f_{\rm incom}$
yields both stable and unstable solutions
of the Potts glass state below $\langle c\rangle_{\rm sp}$,
confirming that the transition to the paramagnetic phase
is discontinuous,
and that $\langle c\rangle_{\rm sp}$
corresponds to the spinodal point.
The Edwards-Anderson order parameter
for both stable and unstable Potts glass states
are also shown in Fig.~\ref{fig:qea},
bearing features similar to those in Fig.~\ref{fig:fincom}.

\begin{figure}[ht]
\centerline{\epsfig{height=7cm,figure=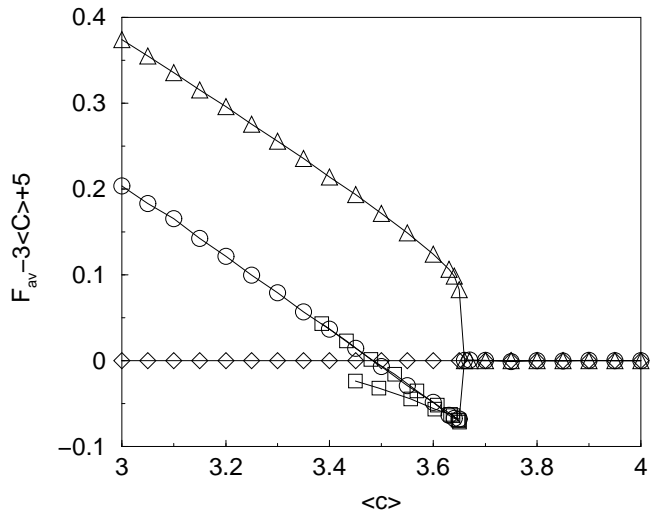}}
\caption{
The dependence of the average free energy on the average connectivity,
after subtracting the baseline $3\langle c\rangle - 5$
of the paramagnetic free energy.
Symbols: $\bigtriangleup$: local estimate of the average energy,
other symbols as in Fig.~\ref{fig:qea}.
Parameters: as in Fig.~\ref{fig:qea}.
\label{fig:fav}}
\end{figure}

Figure~\ref{fig:fav} shows the average free energy.
The paramagnetic free energy of 3$\langle c\rangle-5$
provides a baseline for comparing the energy and free energy
of the different phases.
Below the spinodal point $\langle c\rangle_{\rm sp}$,
the paramagnetic state continues to exist.
It is not accessible by the population dynamics,
but one can find the paramagnetic free energy by first
finding a paramagnetic state
at $\langle c\rangle \ge \langle c\rangle_{\rm sp}$,
and then gradually reducing the connectivity to the desired value.
The resultant paramagnetic free energy
is identical to that found directly in subsection~\ref{subsec:para}.

As shown in Fig.~\ref{fig:fav}, the Potts glass free energy
becomes lower than the paramagnetic free energy near the spinodal
point $\langle c\rangle_{\rm sp}$.
A first order transition
appears to take place at $\langle c\rangle_{\rm c,zic} = 3.48$,
where the free energies of the two states cross each other.
The subscript zic refers to the zero initial condition used here,
as distinguished from the random initial condition
(subscript ric) to be discussed in the next subsection.
However, since the Potts glass energy equals the free energy at zero
temperature, this implies that the average energy is below the
lowest possible energy of $3\langle c\rangle-5$ in the range
$3.48<\langle c\rangle<3.65$! Similar observations of
contradictory results have been observed in the RS ansatz of the
original graph colouring problem~\cite{vanmourik2002,mulet2002}
and the 3-SAT problem~\cite{monasson1997}, This indicates that the
RS ansatz in the present analysis is insufficient, and has to be
improved by including further steps of replica symmetry-breaking.
Furthermore, the solution of the population dynamics is
insensitive to this transition point in the large $N$ limit.
Instead, it yields the Potts glass state above this transition
point right up to the spinodal point $\langle c\rangle_{\rm sp}$.
(For smaller values of $N$, say, $N=1000$, the discontinuous
transition takes place below the spinodal point.) Thus, the
transition at $\langle c\rangle_{\rm sp}$ looks like a zeroth
order one, with a discontinuous jump of the average free energy
from the Potts glass phase below $\langle c\rangle_{\rm sp}$ to
the paramagnetic phase above $\langle c\rangle_{\rm sp}$.

As mentioned in subsection~\ref{subsec:esfinite},
the local and global estimates of the average energy
are different and are given by
Eqs.~(\ref{eq:locale}) and (\ref{eq:globale}) respectively.
The global estimate yields results
identical to the average free energy,
showing that memories about initial conditions
in both variables have been compensated.
However, we observe that the global average energy
is numerically unstable in the Potts glass phase.
For $N=1000$, it diverges from the average free energy
after about 100 steps in the population dynamics.

As shown in Fig.~\ref{fig:fav},
the local estimate of the average energy
is indistinguishable from the global estimate
in the paramagnetic phase.
However, the local estimate is significantly higher
than the global estimate in the Potts glass phase.
Unlike the global estimate
which contradicts the lowest possible energy,
the local estimate remains above it.

\begin{figure}[ht]
\centerline{\epsfig{height=7cm,figure=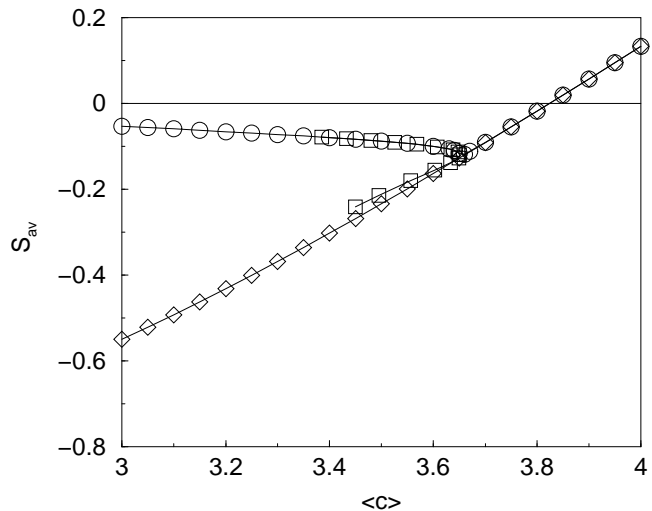}}
\caption{
The dependence of the entropy $S_{\rm av}$ on the average connectivity.
Symbols and parameters: as in Fig.~\ref{fig:qea}.
\label{fig:ent}}
\end{figure}

Next, we consider the entropy.
The entropy of the paramagnetic state
obtained from the theoretical prediction of Eq.~(\ref{eq:sav})
agrees well with the results of population dynamics.
As shown in Fig.~\ref{fig:ent},
the entropy of the paramagnetic state becomes negative
for $\langle c\rangle < \langle c\rangle_{\rm s} = 3.82$,
while the entropy of the Potts glass state is negative throughout.
At the spinodal point $\langle c\rangle_{\rm sp}$, the
entropy exhibits a small discontinuous jump. Clearly, results for
$\langle c\rangle < \langle c\rangle_{\rm sp}$
should be investigated using a replica symmetry-breaking ansatz
to identify the exact transition point,
which is beyond the scope of this paper.

\subsection{Initial conditions}
\begin{figure}[ht]
\begin{center}
\begin{picture}(380,360)
\put(12,170){\epsfxsize=68mm  \epsfbox{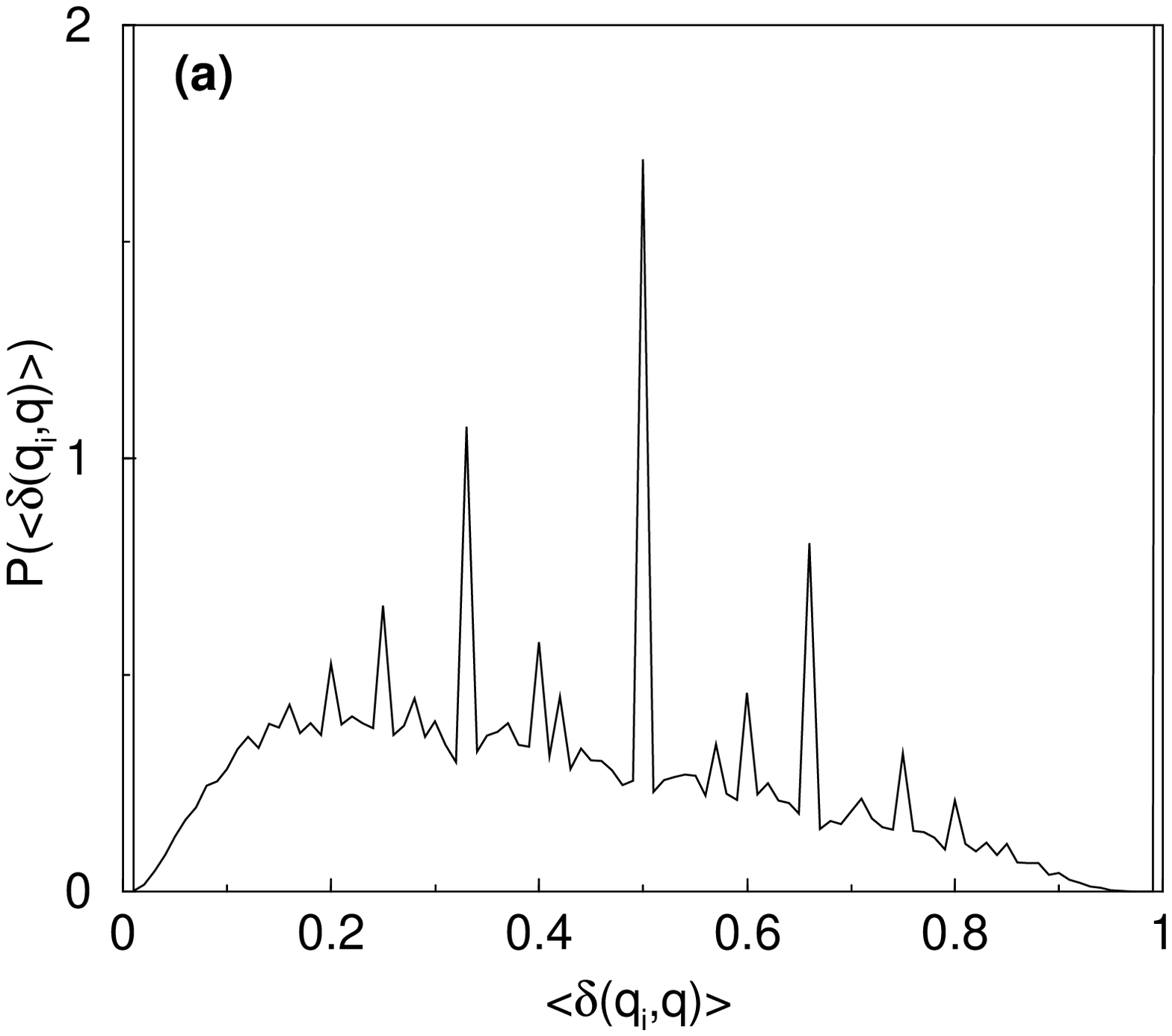}}
\put(210,170){\epsfxsize=70mm  \epsfbox{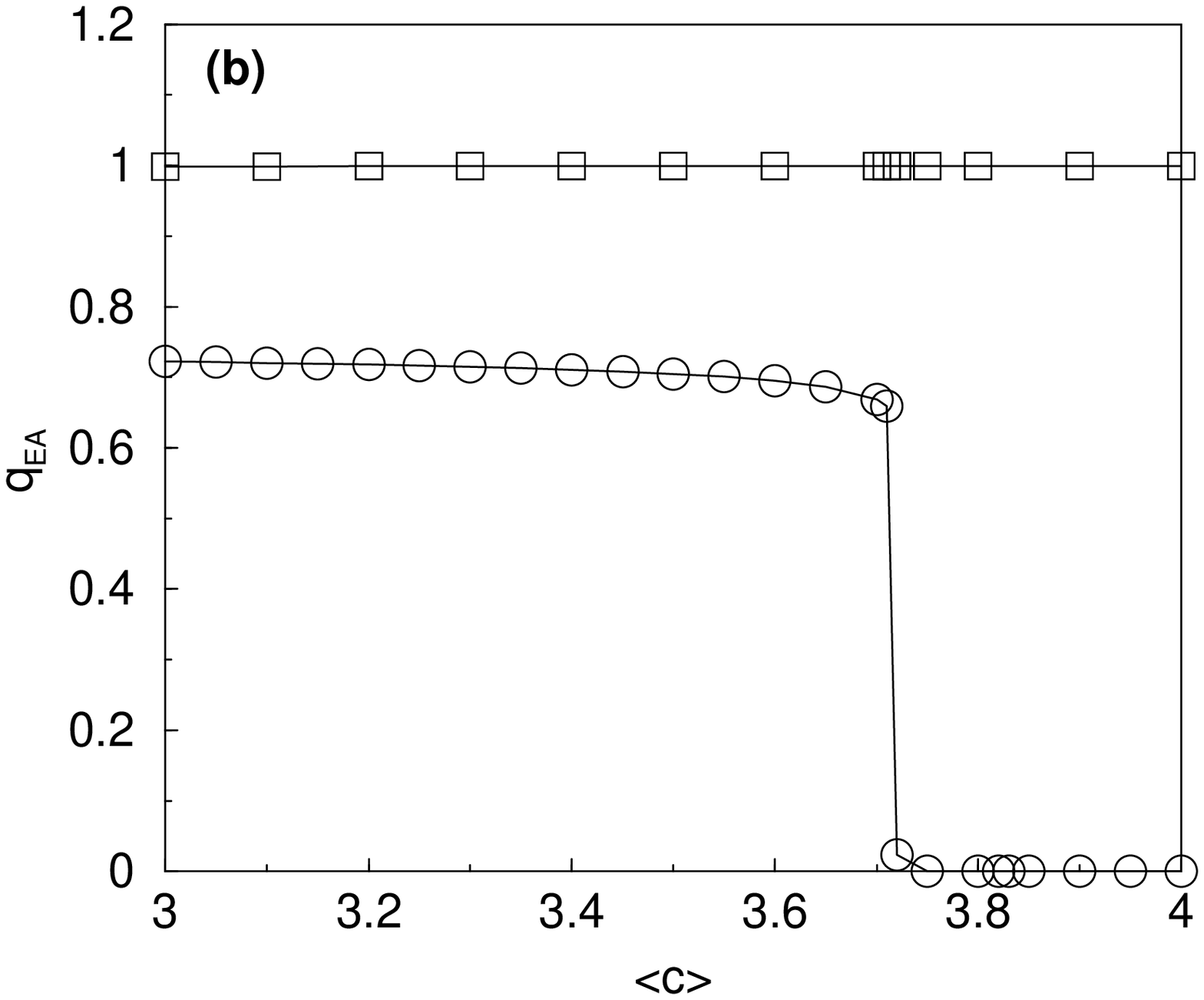}}
\put(0,0){\epsfxsize=73mm  \epsfbox{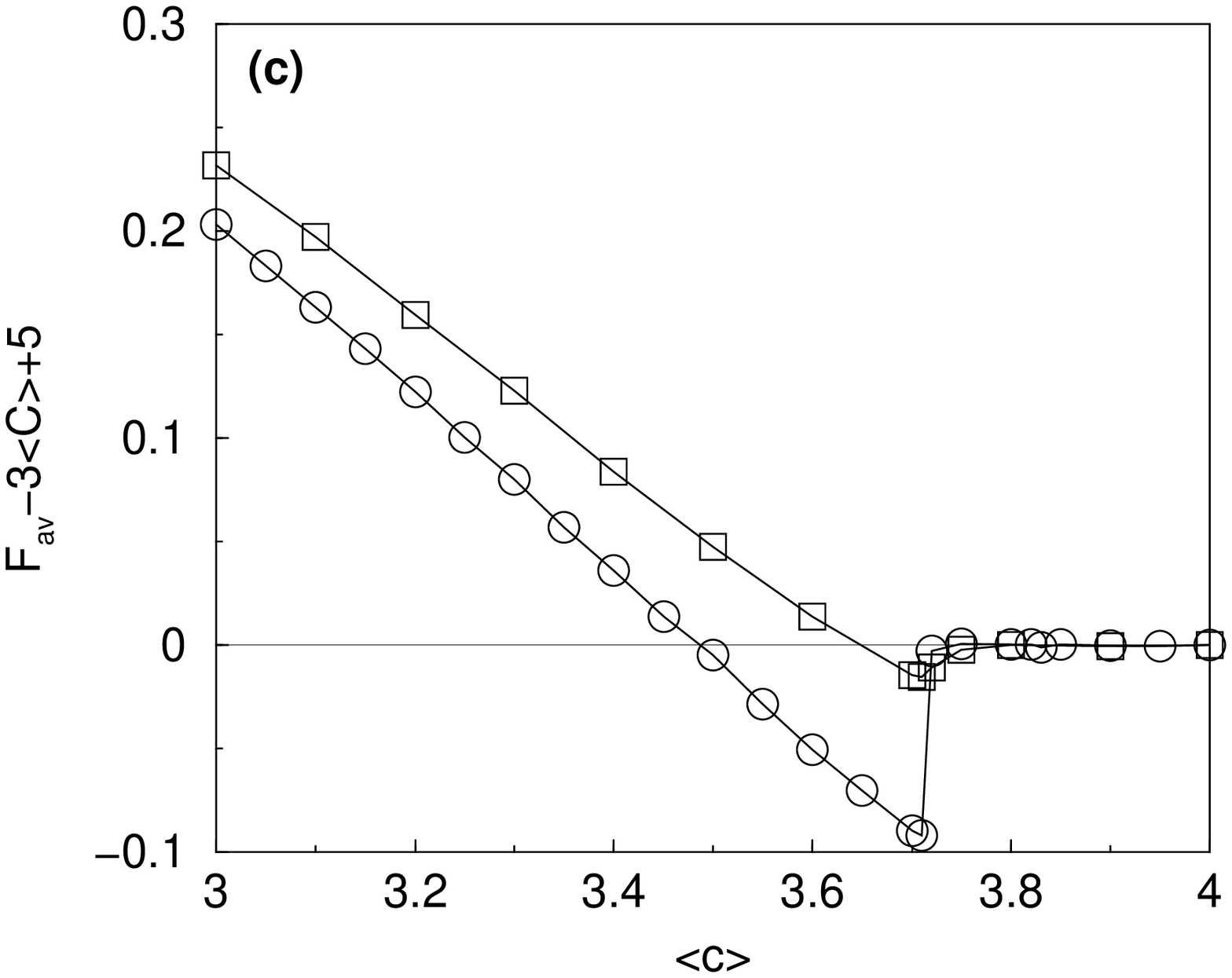}}
\put(210,0){\epsfxsize=70mm  \epsfbox{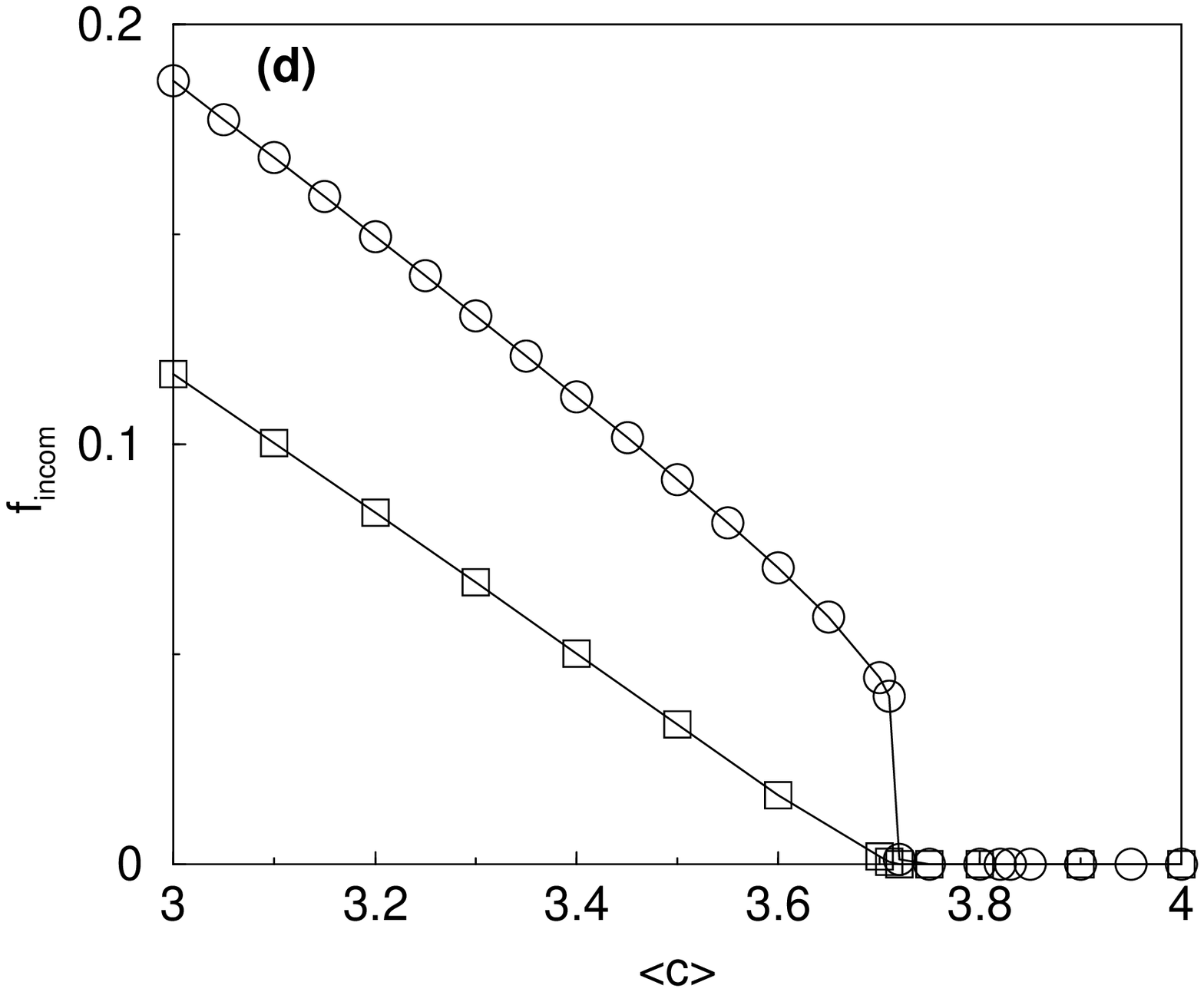}}
\end{picture}
\caption{
Results for system size $N\!=\!10000$, $Q = 4$ and 30
samples, obtained from the steady state solution of the population
dynamics using zero and random initial conditions
(labelled $\bigcirc$ and $\square$ respectively).
(a) The colour moments distribution
obtained from the zero initial condition at $\langle c\rangle=3$.
(b) The Edwards-Anderson order parameter $q_{\rm EA}$.
(c) The average free energy after subtracting
the baseline $3\langle c\rangle-5$ of the paramagnetic free
energy.
(d) The incomplete fraction.
\label{fig:ric}}
\end{center}
\end{figure}

One puzzle of our results is that
the Edwards-Anderson order parameter remains at a level
around 0.7 in the entire Potts glass phase.
This implies that a considerable fraction of nodes
have free choices of colours even in the Potts glass phase.
This is illustrated by the distribution of colour moments
$\langle\delta(q_i,q)\rangle$ in Fig.~\ref{fig:ric}(a),
which consists of a continuous background
with peaks at simple rational numbers
(1/5, 1/4, 1/3, 2/5 etc.).
In fact, the existence of free spins at zero temperature
has been considered an indication of broken replica
symmetry~\cite{mulet2002}.

However, this is apparently inconsistent with extrapolations from
finite temperatures, which will be discussed in the next section.
As will be seen, $q_{\rm EA}$ approaches 1 in the limit of low but
finite temperature, implying that all nodes lose the freedom of
choosing more than one colour.

To resolve this inconsistency,
we consider the effects of introducing a small randomness
in the initial condition, that is,
a small random bias is added to the initial values
of the vertex free energies,
which take integer values otherwise.
Such randomness were known to cause significant changes
in the optimal solution in the graph bipartitioning problem,
where the field distribution is initialised
to a rectangular distribution~\cite{wong1988}.

Figure~\ref{fig:ric}(b) shows that when a very small randomness is
introduced in the initial condition, the final values of the
Edwards-Anderson order parameter $q_{\rm EA}$ remain around 1 in both
the paramagnetic and Potts glass phase. This means that effectively all
spins are frozen due to the randomness in the initial condition. 
The distribution of colour moments
consists of two delta function peaks,
located at $\langle\delta(q_i,q)\rangle=0$ and $1$ respectively.
This is consistent with the extrapolation
of finite temperature results.
The difference between zero temperature
and low but finite temperature distributions
was also observed in the RS approximation
of the original graph colouring
problem~\cite{mulet2002,vanmourik2002}.

Randomness in the initial condition causes a significant change in
the transition point between the Potts glass and paramagnetic states.
Figure~\ref{fig:ric}(c) shows that the average free energy of the
Potts glass state crosses that of the paramagnetic state
at $\langle c\rangle_{\rm c,zic}=3.48$ and
$\langle c\rangle_{\rm c,ric}=3.65$
for the zero and random initial conditions, respectively.
As far as we can tell from our numerical precision,
$\langle c\rangle_{\rm c,ric}=3.65$
is effectively the same as the spinodal point
$\langle c\rangle_{\rm sp}=3.65$.
As will be seen in the next section,
the transition point $\langle c\rangle_{\rm c,ric}$
is consistent with the phase transition line at finite temperatures.

The effects of randomness in the initial condition on the
performance are shown in Fig.~\ref{fig:ric}(d).
For the random initial condition,
the incomplete fraction in the Potts glass phase
vanishes effectively continuously to 0 at $\langle c\rangle_{\rm sp}$.
This is in contrast with the incomplete fraction
for the zero initial condition,
which is much higher, and vanishes discontinuously at the spinodal point.

The entropy is effectively zero in both the Potts glass phase
and the paramagnetic phase in the case of random initial conditions.
This is different from the case of zero initial conditions
shown in Fig.~\ref{fig:ent},
in which the entropy is negative
in the entire Potts glass phase and part of the paramagnetic phase.

\subsection{Evolution of damages}
\label{subsec:damage}
To illustrate the difference
between the paramagnetic and Potts glass phases,
we consider the evolution of damages for different average
connectivities $\langle c\rangle $.
The damaged configuration, with colours
$\{q_i'\}$, is initialised identically to $\{q_i\}$,
except that the colours of the descendants of
one randomly chosen node $j$ have been inverted, that is, $q_{k}=Q
- q_{k}'$ where $k$ are the descendants of node $j$. We define the
\textit{distance measure} between {\{}$q_{i}${\}} and
{\{}$q_{i}$'{\}} as the distance between the colour moments
\begin{equation}
\label{eq:distance} d=\frac{1}{N}\sum\limits_i
    {\sum\limits_{q=1}^Q {\left( {\left\langle \delta (q_i ,q)
    \right\rangle -\left\langle \delta (q_i ',q) \right\rangle }
    \right)^2} } .
\end{equation}

\begin{figure}[htbp]
\centerline{\epsfig{height=7cm,file=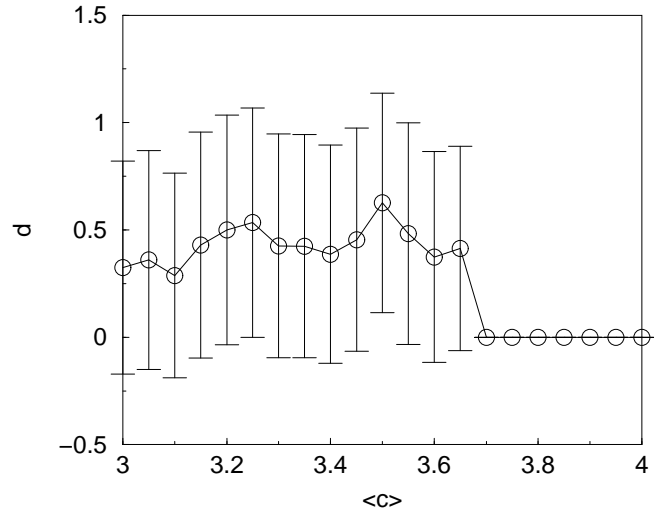}}
\caption{
The dependence of the distance measure $d$ on the average
connectivity $\langle c\rangle $ using population dynamics with
10000 nodes, $Q=4$ and 30 samples.
\label{fig:damage}}
\end{figure}

We monitor the population dynamics of the colour configuration
{\{}$q_{i}${\}} and its \textit{damaged} configuration
{\{}$q_{i}$'{\}}. They evolve with the same sequence of updates
and choice of descendants. As shown in Fig.~\ref{fig:damage}, the
distance is nonzero in the Potts glass phase, but vanishes in the
paramagnetic phase. This shows that multiple solutions of the
saddle point equation exist in the Potts glass phase, but the solution
is unique in the paramagnetic phase.
The spread of damage is consistent
with the instability of the replica symmetric
solution in the Potts glass phase.

\section{Finite Temperature Behaviour}
\label{sec:finitetemp}

\subsection{The example of $\langle c\rangle=3$}
\begin{figure}[ht]
\begin{center}
\begin{picture}(380,180)
\put(0,0){\epsfxsize=70mm  \epsfbox{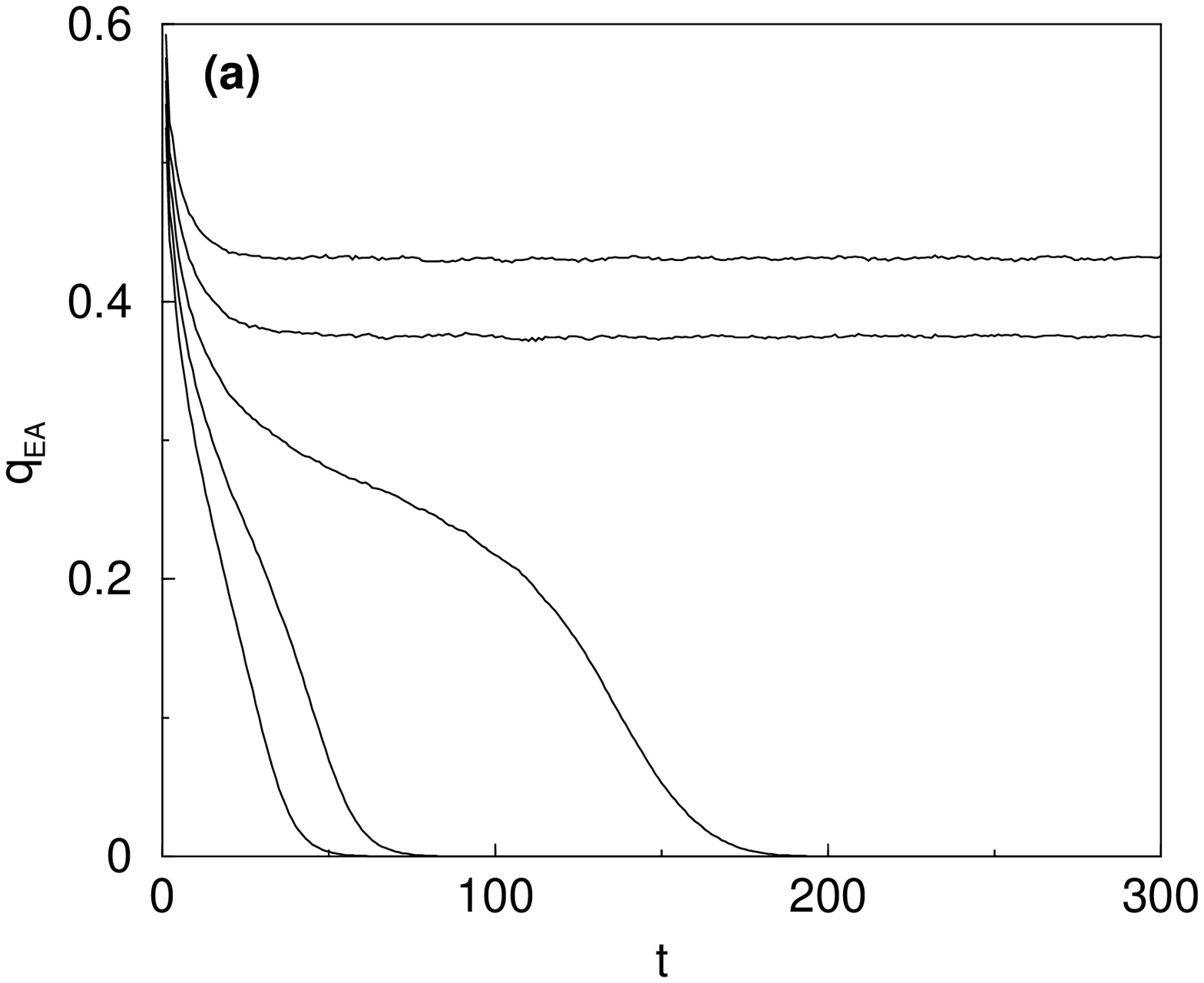}}
\put(200,0){\epsfxsize=70mm \epsfbox{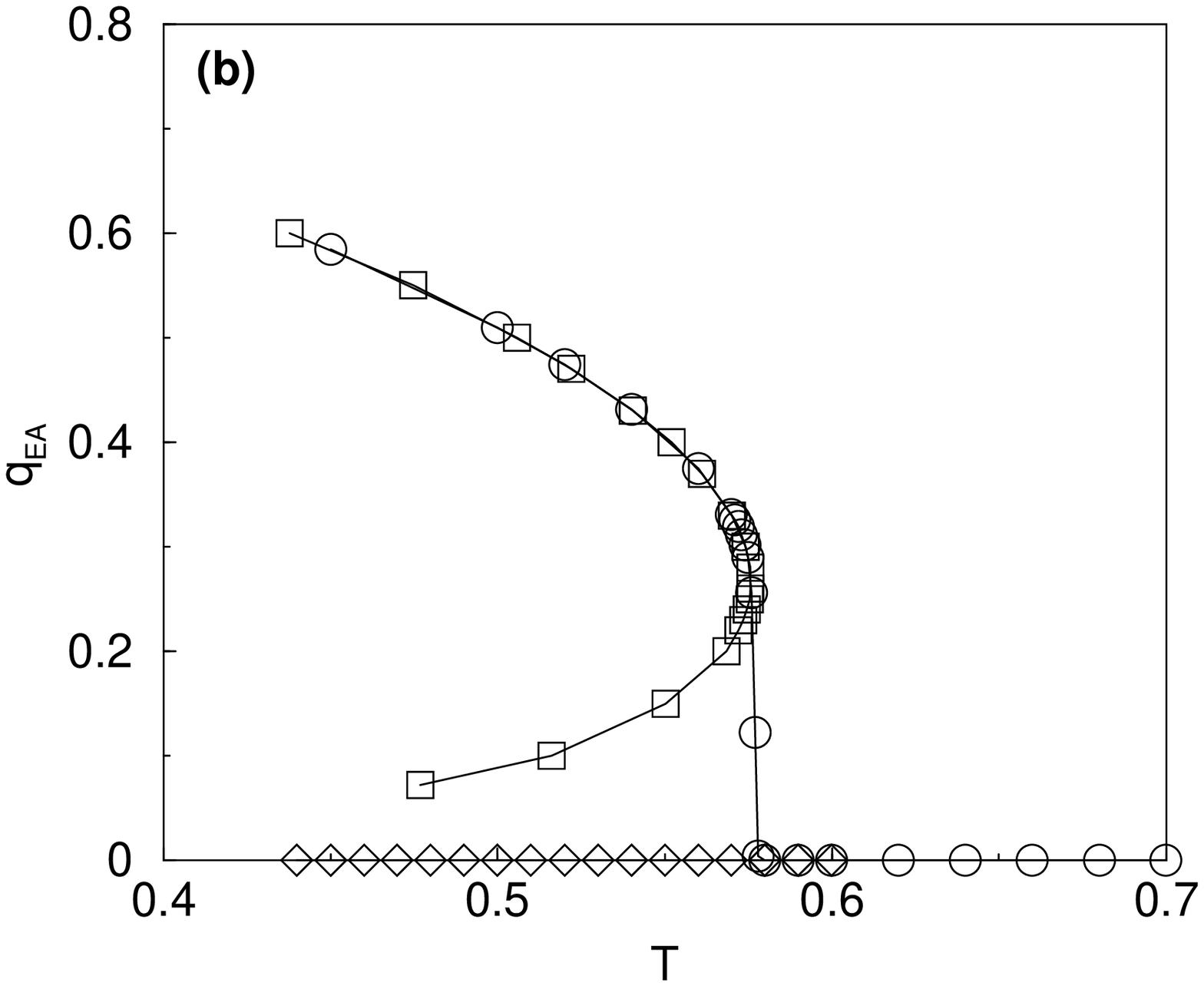}}
\end{picture}
\caption{
(a) The evolution of the Edwards-Anderson order parameter $q_{\rm EA}$
in the population dynamics at $\langle c\rangle=3$
and $T=0.54, 0.56, 0.58, 0.60, 0.62$ (top to bottom).
(b)The dependence of $q_{\rm EA}$ at the steady state
on temperature $T$.
Symbols: thermodynamic state ($\bigcirc$),
Potts glass state ($\square$),
paramagnetic state ($\lozenge$).
Parameters: $N=10000$, $Q=4$ and 30 samples.
\label{fig:qea3}}
\end{center}
\end{figure}

Further insights about the thermodynamic behaviour
can be obtained by considering the finite temperature behaviour.
Let us first study the example of $\langle c\rangle=3$.
Figure~\ref{fig:qea3}(a) shows that $q_{\rm EA}$
of the thermodynamic state vanishes
at temperatures above 0.575.
To verify that this phase transition is discontinuous,
we look for solutions of the population dynamics
with variable $T$ for given values of $q_{\rm EA}$,
which yield the Potts glass state.
As shown in Fig.~\ref{fig:qea3}(b),
the Potts glass phase with positive $q_{\rm EA}$
does not vanish continuously into the paramagnetic phase.
Rather, its stable and unstable branches
merge at the temperature 0.575,
which is therefore identified to be the spinodal temperature.

\begin{figure}[ht]
\begin{center}
\begin{picture}(380,180)
\put(0,0){\epsfxsize=70mm  \epsfbox{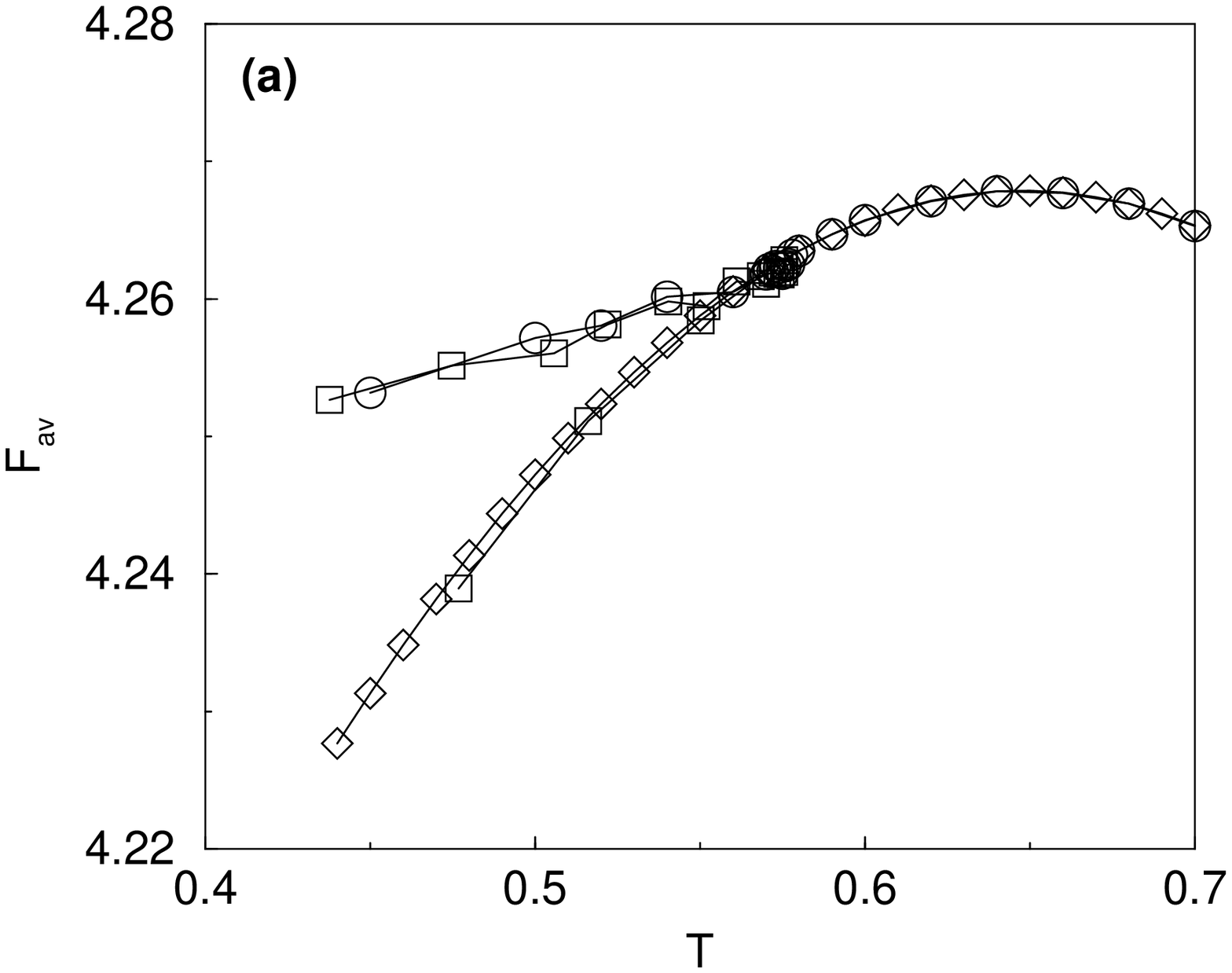}}
\put(200,0){\epsfxsize=70mm \epsfbox{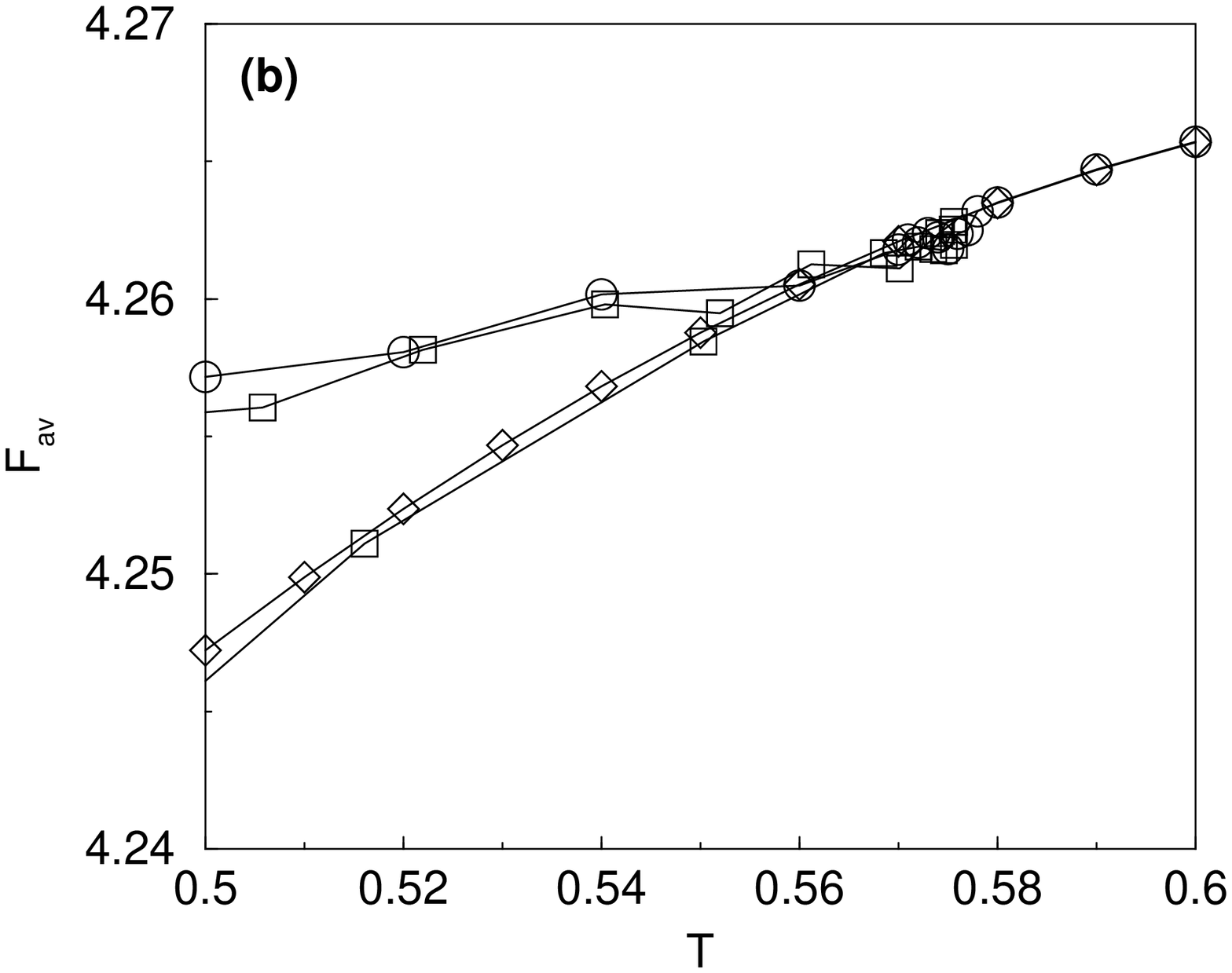}}
\end{picture}
\caption{
The dependence of the average free energy $F_{\rm av}$
on temperature at $\langle c\rangle=3$.
Symbols and parameters: as in Fig.~\ref{fig:qea3}(b).
\label{fig:fav3}}
\end{center}
\end{figure}

Figure~\ref{fig:fav3}(a) shows the free energies
of the paramagnetic state
and the results of the population dynamics.
The free energy at the paramagnetic state
reaches a maximum at $T=0.65$.
Below this temperature, the entropy becomes negative.
The population dynamics is in good agreement
with the paramagnetic state
down to the spinodal temperature,
below which the population dynamics
deviates from the paramagnetic state.

Figure~\ref{fig:fav3}(b) shows the free energies
in the neighbourhood of the spinodal temperature,
including the stable and unstable branches
of the Potts glass state.
The free energies of the Potts glass and paramagnetic states
become equal at $T=0.56$.
While this can be interpreted
as the thermodynamic transition temperature,
we observe that it is not relevant to the population dynamics,
in which the jump of $q_{\rm EA}$,
as shown in Figs.~\ref{fig:qea3}(a) and (b),
takes place at the spinodal temperature instead.
This behaviour is consistent
with the irrelevance of the first order transition point
$\langle c\rangle_{\rm c,zic}=3.48$ at zero temperature,
as described in subsection~\ref{subsec:para-glass}.

\begin{figure}[ht]
\begin{center}
\begin{picture}(380,180)
\put(0,0){\epsfxsize=70mm  \epsfbox{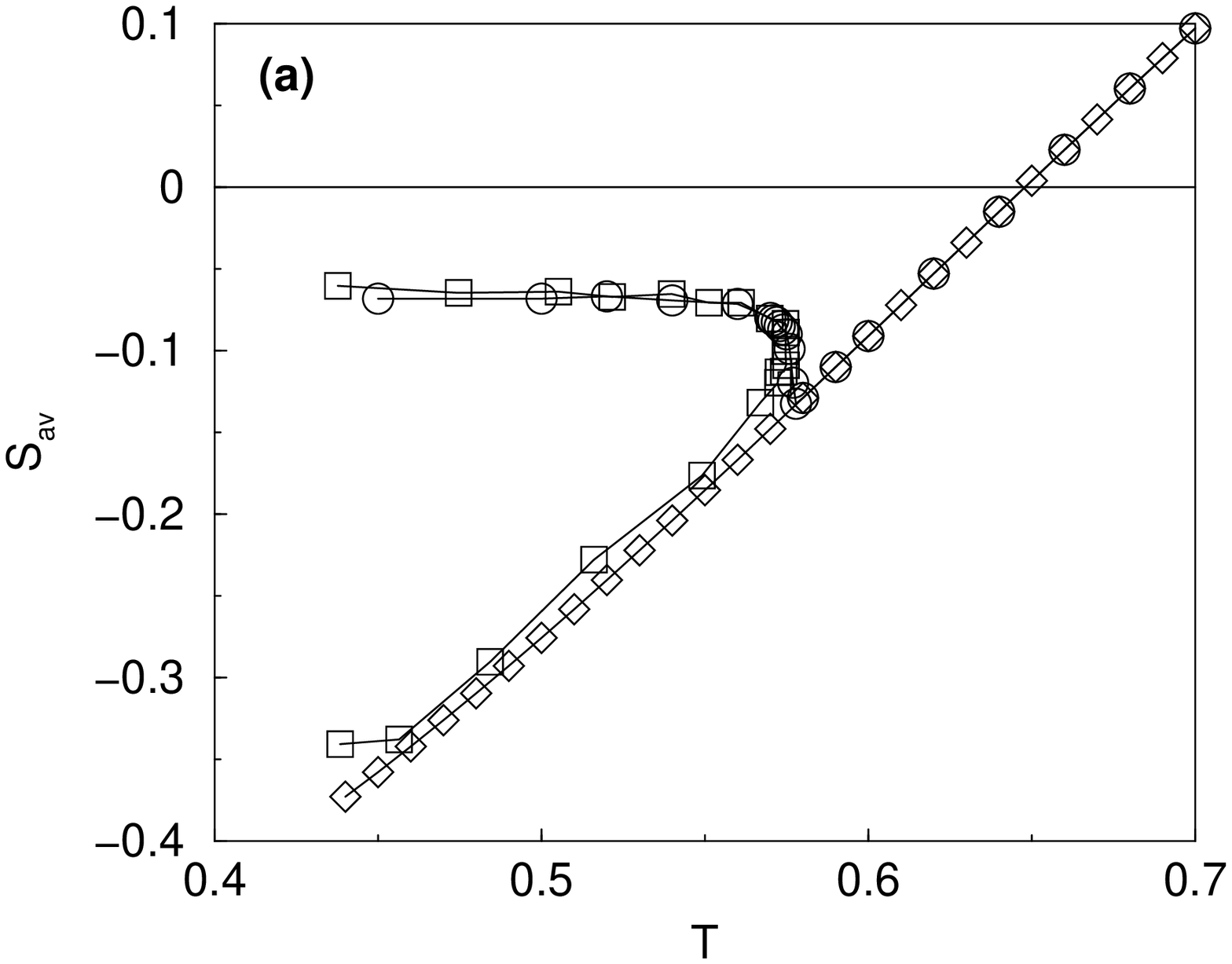}}
\put(200,0){\epsfxsize=70mm \epsfbox{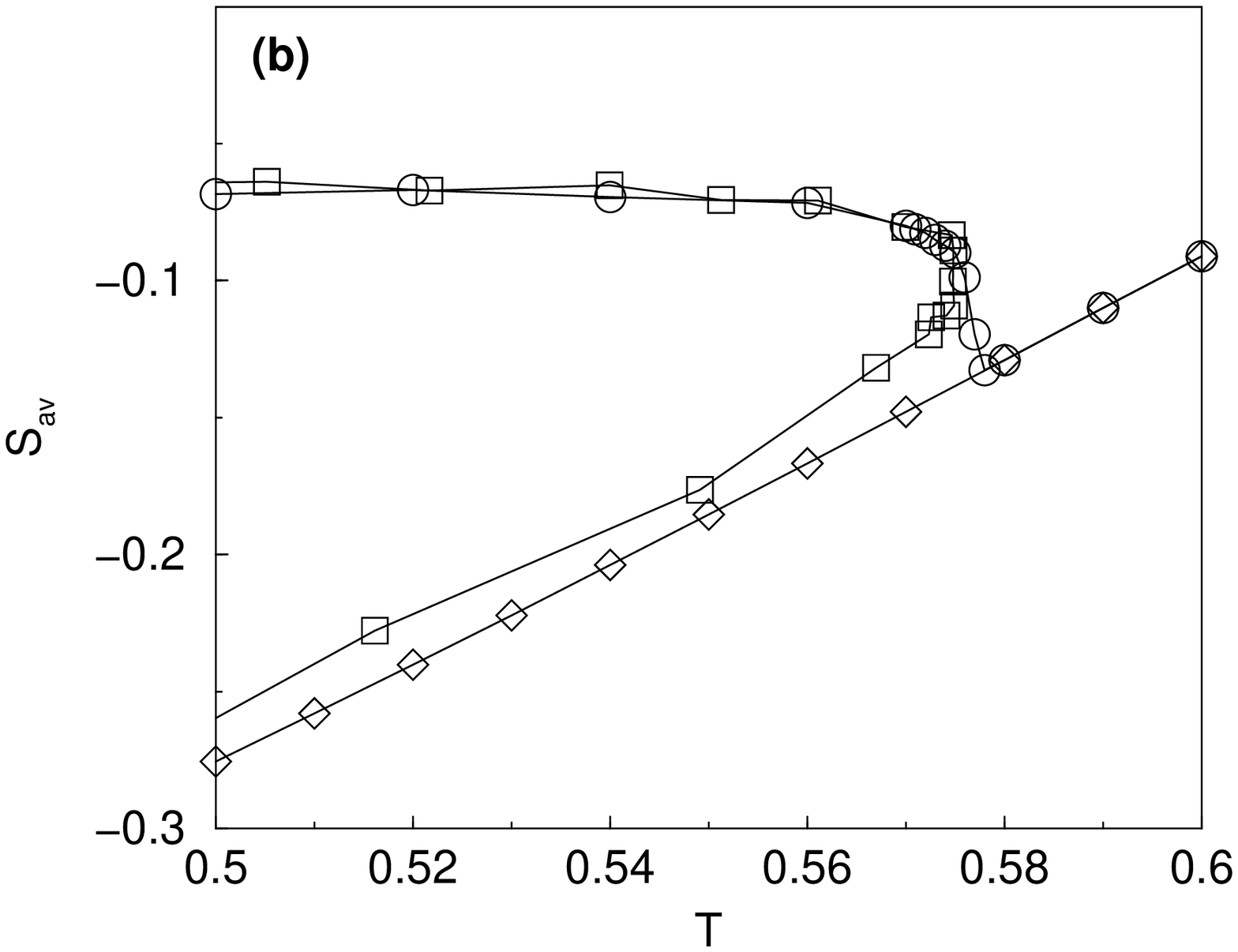}}
\end{picture}
\caption{
The dependence of the average entropy $S_{\rm av}$
on temperature at $\langle c\rangle=3$.
Symbols and parameters: as in Fig.~\ref{fig:qea3}(b),
except that $N=1000$ and 100 samples for the Potts glass state.
\label{fig:ent3}}
\end{center}
\end{figure}

The behaviour of the entropy
is shown in Fig.~\ref{fig:ent3}(a).
The entropy of the paramagnetic state
becomes negative below $T=0.65$.
The stable and unstable branches of the Potts glass state
are shown in Fig.~\ref{fig:ent3}(b),
and the population dynamics yields results
jumping discontinuously
from the stable branch of the Potts glass state
to the paramagnetic state at the spinodal temperature.

Regions of negative entropy are often found in spin glasses.
They usually signal that the RS ansatz is unstable.
However, in the original Sherrington-Kirkpatrick model,
the region of negative entropy is restricted
to the low temperature regime
deep inside the spin glass phase~\cite{sherrington1975,kirkpatrick1978}.
In contrast, the region of negative entropy at $\langle c\rangle=3$
spans the entire Potts glass phase
and even covers part of the paramagnetic phase.
This indicates that frustration effects
in the present model is unusually strong.

\begin{figure}[ht]
\centerline{\epsfig{height=7cm,figure=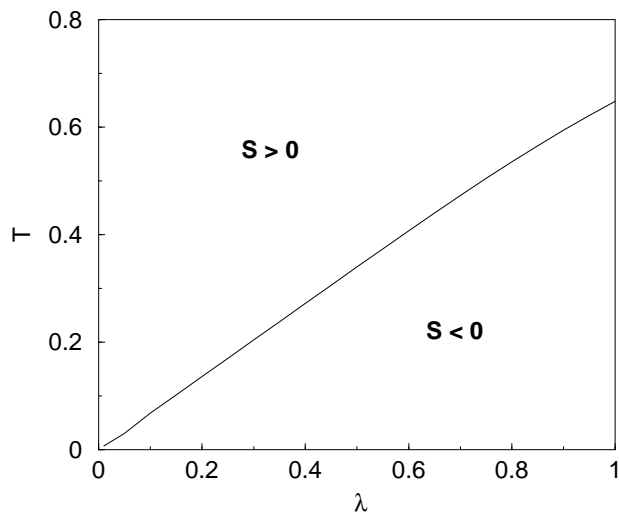}}
\caption{
Regions of positive and negative entropies
of the paramagnetic state for $\langle c\rangle=3$ and $Q=4$.
\label{fig:ent_pm}}
\end{figure}

We propose that this increased frustration effect
is a consequence of the second nearest neighbouring interactions
present in the colour diversity problem,
and does not exist in most models investigated so far.
To verify this, we consider the model
\begin{equation}
        E=\sum_i\left[4+2\sum_{j\in N_i}\delta(q_i,q_j)
        +2\lambda\sum_{j\ne k\in N_i}\delta(q_j,q_k)\right].
\end{equation}
The cases $\lambda=0$ and 1 correspond to the graph colouring
and colour diversity problems respectively,
We will consider the range $0\le\lambda\le 1$.
In the paramagnetic phase, expressions for the entropy
can be derived analogously to Appendix B.
As shown in Fig.~\ref{fig:ent_pm},
the region of negative entropy of the paramagnetic state
shrinks when the second nearest neighbouring interaction is reduced.
Thus, in the absence of second nearest neighbouring interaction,
the region of paramagnetic phase with negative entropy
is preempted by the Potts glass phase.

\subsection{General values of $\langle c\rangle$}
\begin{figure}[htbp]
\centerline{\epsfig{height=7cm,file=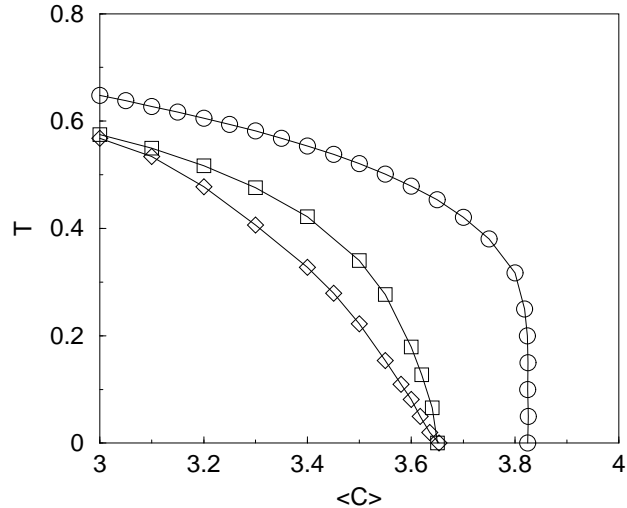}}
\caption{
The zero entropy line ($\bigcirc$),
spinodal line ($\square$) and the paramagnetic-glass transition line
($\lozenge$) in the space of the average connectivity
$\langle c\rangle $ and temperature $T$ for $Q=4$.
\label{fig:transition}}
\end{figure}

For general values of $\langle c\rangle$
we will consider three transition lines
in the space of $\langle c\rangle$ and $T$:
the zero entropy line
in the paramagnetic phase, the spinodal line of the glassy state,
and the paramagnetic-glass transition line.
The transition lines are plotted in Fig.~\ref{fig:transition}.
When extrapolated to $T = 0$, the zero entropy, spinodal and
free-energy crossing lines pass through the points $\langle
c\rangle = 3.82$, $3.65$ and $3.65$, respectively, in full
agreement with the results obtained for the zero temperature case.

In summary, the system has a paramagnetic phase at high
temperature or high connectivity. Inferring from the studies of
the graph colouring problem~\cite{mulet2002,braunstein2003},
we expect that a phase transition to
replica symmetry-breaking states takes place at the high
temperature (and high connectivity) side of the zero entropy line,
even when the system is still in the paramagnetic state. However,
the location of this transition cannot be found in the present
framework of replica symmetry.

Nevertheless, the replica symmetric solution
has provided us insights on the full solution,
suggesting the following picture.
One expects the existence
of the spinodal line, where the Potts glass state with a nonzero
Edwards-Anderson order parameter exists in its low temperature
(and low connectivity) side. The Potts glass state exists as a
metastable state in the vicinity of the spinodal line. Then, at
the low temperature (and low connectivity) side of the
paramagnetic-glass transition line, the Potts glass state becomes
thermodynamically stable.

\section{Conclusion}

We have studied the macroscopic behaviour in the colour diversity
problem, a variant of the graph colouring problem of significant
practical relevance, especially in the area of distributed storage
and content distribution.
To cope with the presence of
second nearest neighbouring interactions,
the analysis makes use of vertex free energies of two arguments,
which enable us to study the behaviour in the RS analysis,
and lays the foundation for future analyses
incorporating replica symmetry-breaking effects.
The analysis is successfully applied
to graphs with mixed connectivities.

For $Q=4$ and graphs with linear connectivity $3\le\langle c\rangle\le 4$,
the RS analysis identifies three transition lines according to:
(1) when the entropy becomes negative
(ending at $\langle c\rangle_{\rm s} = 3.82$ when $T=0$),
signalling the breakdown of the RS ansatz;
(2) when $q_{\rm EA}$ becomes multiple-valued function of $T$
-- the spinodal point
(ending at $\langle c\rangle_{\rm sp} = 3.65$ when $T=0$); and
(3) the free-energy crossing point
between the paramagnetic and Potts glass state
(ending at $\langle c\rangle_{\rm c} = 3.65$ when $T$ approaches 0).
The regime of negative entropy is so extensive
that it covers the entire Potts glass phase
as well as part of the paramagnetic phase,
and can be attributed to the increased frustration
due to the presence of second nearest neighbouring interactions.

The picture that emerges is that the system is in a paramagnetic
state at high temperature or high connectivity; the RS ansatz
breaks down prior to the temperature that identifies the zero
entropy transition point. The Potts glass state exists first as a
metastable state but becomes dominant at a lower temperature
(connectivity).
Evidence from the population dynamics
shows that the discontinuous transition takes place
at the spinodal point rather than the crossing point.
However, the RS analysis results in the average energy
falling below the lowest possible energy
for $3.48<\langle c\rangle<3.65$,
and a region of negative entropy.

Since the entropy remains positive
at the colourable-uncolourable
transition~\cite{mulet2002,braunstein2003},
we conjecture that if replica symmetry-breaking
is taken into account,
the Potts glass-paramagnetic transition should take place
at the higher temperature (and high connectivity) side
of the zero entropy line.
For the optimisation of the colour diversity,
one should consider $T=0$,
implying that the incomplete-complete transition
should take place at $\langle c\rangle$
beyond $\langle c\rangle_s=3.82$.
This estimate of the transition point
seems to be supported by simulation results
using the Walksat and BP algorithms~\cite{bounkong2006}.

In summary, we have demonstrated the value of different analytical
approaches and the use of population dynamics  in elucidating the
system behaviour of the colour diversity problem on a sparse
graph. They provide insights on
the estimates of the transition points,
the existence of metastable states, and the nature of phase
transitions.

\subsubsection*{Acknowledgements}
We thank Lenka Zdeborov\'a, David Sherrington,
Bill Yeung, Edmund Chiang for meaningful discussions,
and Stephan Mertens for drawing our attention to \cite{mccormick1983}.
This work is partially supported by research grants DAG04/05.SC25,
DAG05/06.SC36, HKUST603606 and HKUST603607
of the Research Grant Council of Hong Kong, by
EVERGROW, IP No. 1935 in the complex systems initiative of the FET
directorate of the IST Priority, EU FP6 and EPSRC grant
EP/E049516/1.

\appendix

\section{Replica Approach to Colour Diversity}

Consider the minimisation of the energy (cost function)
on a graph of connectivity $c$:
\begin{equation}
    E=\sum_i\sum_{j_1\ne\cdots\ne j_c}
    a_{ij_1}\cdots a_{ij_c}
    \phi(q_i,q_{j_1},\cdots,q_{j_c}),
\end{equation}
where $\phi$ is symmetric with respect to
the permutation of the neighbours,
$q_i\in\{1,\cdots,Q\}$,
and $a_{ij}=1$
if nodes $i$ and $j$ are connected on the graph,
and 0 otherwise.
Since there are $Q^{c+1}$ values of the function $\phi$,
one can write
\begin{equation}
    \phi(q_i,q_{j_1},\cdots,q_{j_c})
    =\sum_{m_0,\cdots,m_c=1}^Q
    \phi_{m_0\cdots m_c}q_i^{m_0}\cdots q_{j_c}^{m_c}.
\end{equation}
The partition function is
\begin{equation}
    Z={\rm Tr}_{\mathbf q}\exp\left[-\beta
    \sum_i\sum_{j_1\ne\cdots\ne j_c}
    a_{ij_1}\cdots a_{ij_c}
    \sum_{\mathbf m}
    \phi_{m_0\cdots m_c}q_i^{m_0}\cdots q_{j_c}^{m_c}
    \right].
\end{equation}
The replicated partition function,
averaged over all graph configurations with connectivity $c$,
is given by
\begin{eqnarray}
    \langle Z^n\rangle
    =&&\frac{1}{\cN}\sum_{a_{ij}=0,1}
    \prod_i\delta\left(\sum_j a_{ij}-c\right)
    {\rm Tr}_{\mathbf q}\exp\left[-\beta
    \sum_i\sum_{j_1\ne\cdots\ne j_c}
    a_{ij_1}\cdots a_{ij_c}\right.
    \nonumber\\
    &&\left.\times\sum_{\mathbf m,\alpha}
    \phi_{\mathbf m}(q_i^\alpha)^{m_0}\cdots(q_{j_c}^\alpha)^{m_c}
    \right],
\end{eqnarray}
where $\cN$ is the total number of graph representations
with connectivity $c$.

It is convenient to express the exponential argument
as an unrestricted sum over the nodes $j_1,\cdots,j_c$,
\begin{eqnarray}
    &&-\frac{\beta}{c!}\sum_i\left(
    \sum_{j_1\cdots j_c}-B_2\sum_{j_1=j_2}\sum_{j_3\cdots j_c}
    -\cdots+(-)^{c-1}B_c\sum_{j_1=\cdots j_c}\right)
    \nonumber\\
    &&\times a_{ij_1}\cdots a_{ij_c}
    \sum_{{\mathbf m},\alpha}
    \phi_{\mathbf m}(q_i^\alpha)^{m_0}\cdots(q_{j_c}^\alpha)^{m_c},
\end{eqnarray}
where $B_2,\cdots,B_c$ are integers accounting
for the over-counting in rewriting the summations in terms of equal indices.
Their precise values are not required in our final result.
This allows us to factorise the expression into
\begin{eqnarray}
    &&-\frac{\beta}{c!}\sum_{{\mathbf m},\alpha}\phi_{\mathbf m}
    \sum_i(q_i^\alpha)^{m_0}\left\{
    \left[\sum_{j_1}a_{ij_1}(q_{j_1}^\alpha)^{m_1}\right]\cdots
    \left[\sum_{j_c}a_{ij_c}(q_{j_c}^\alpha)^{m_c}\right]\right.
    \nonumber\\
    &&-B_2\left[\sum_{j_1}a_{ij_1}(q_{j_1}^\alpha)^{m_1+m_2}\right]
    \left[\sum_{j_c}a_{ij_3}(q_{j_3}^\alpha)^{m_3}\right]
    \cdots
    \left[\sum_{j_c}a_{ij_c}(q_{j_c}^\alpha)^{m_c}\right]
    \nonumber\\
    &&\left.+\cdots+(-)^{c-1}B_c
    \left[\sum_{j_1}a_{ij_1}(q_{j_1}^\alpha)^{m_1+\cdots+m_c}\right]
    \right\}.
\label{eq:factorsum}
\end{eqnarray}
Following steps similar to those in \cite{wong2007}, one gets
\mathindent = 1 pc
\begin{eqnarray}
    \langle Z^n\rangle
    &=&\exp N\left\{\frac{c}{2}
    -c\sum_{{\mathbf r},{\mathbf s}}
    \hat Q_{{\mathbf r},{\mathbf s}}Q_{{\mathbf r},{\mathbf s}}
    +\ln{\rm Tr}_{\mathbf q}\prod_{m,\alpha}\left(
    \int\frac{d\hat h_m^\alpha dh_m^\alpha}{2\pi}
    \exp\left[\sum_{m,\alpha}\left(i\hat h_m^\alpha h_m^\alpha\right)
    \right]\right)\right.
    \nonumber\\
    &&\times\left[\sum_{r_m^\alpha,s_m^\alpha}
    \hat Q_{{\mathbf r},{\mathbf s}}\prod_{m,\alpha}
    (-i\hat h_m^\alpha)^{r_m^\alpha}(q^\alpha)^{ms_m^\alpha}
    +\frac{1}{2}\sum_{r_m^\alpha,s_m^\alpha}\prod_{m,\alpha}
    \frac{(-i\hat h_m^\alpha)^{s_m^\alpha}}{r_m^\alpha!s_m^\alpha!}
    (q^\alpha)^{mr_m^\alpha}\right]^c
    \nonumber\\
    &&\times\exp\left\{-\frac{\beta}{c!}\sum_{{\mathbf m},\alpha}
    \phi_{\mathbf m}(q^\alpha)^{m_0}\left[
    h_{m_1}^\alpha\cdots h_{m_c}^\alpha
    -B_2 h_{m_1+m_2}^\alpha h_{m_3}^\alpha\cdots h_{m_c}^\alpha
    +\cdots\right.\right.
    \nonumber\\
    &&\left.+(-)^{c-1}B_c h_{m_1+\cdots+m_c}^\alpha
    \right]\Biggr\}\Biggr\},
\label{eq:partn}
\end{eqnarray}
where $Q_{{\mathbf r},{\mathbf s}}$ and
$\hat Q_{{\mathbf r},{\mathbf s}}$ are given by
the saddle point equations of Eq.~(\ref{eq:partn}).

Consider the generating function
\begin{equation}
    P_{\mathbf s}({\mathbf z})
    =\sum_{\mathbf r}
    Q_{{\mathbf r},{\mathbf s}}
        \prod_{m,\alpha}\frac{(z_\alpha)^{mr_m^\alpha}}{r_m^\alpha!}.
\label{eq:gen}
\end{equation}
In the replica symmetric ansatz,
we consider functions of the form
\begin{equation}
    P_{\mathbf s}({\mathbf z})
        =\left\langle\prod_\alpha
        \left({\rm Tr}_{\bmu} R(z_\alpha,\mu^\alpha|{\mathbf T})
        (\mu^\alpha)^{\sum_m ms_m^\alpha}\right)\right\rangle.
\end{equation}
Substituting the saddle point equation
for $Q_{{\mathbf r},{\mathbf s}}$ into Eq.~(\ref{eq:gen}),
one finds $P_{\mathbf s}({\mathbf z})=\sN_P/\sD_P$ where
\begin{eqnarray}
    \sN_P&=&\Biggl\langle\prod_\alpha\Biggl\{{\rm Tr}_q
    \prod_{k=1}^{c-1}\left[{\rm Tr}_{\bmu_k}
    R(q^\alpha,\mu_k^\alpha|{\mathbf T}_k)\right]
    \prod_m(q^\alpha)^{ms_m^\alpha}
    \nonumber\\
    &&\times\exp\Biggl[-\frac{\beta}{c!}\sum_{{\mathbf m},\alpha}
    \phi_{\mathbf m}(q^\alpha)^{m_0}\Biggl(
    h_{m_1}^\alpha\cdots h_{m_c}^\alpha
    -B_2 h_{m_1+m_2}^\alpha h_{m_3}^\alpha\cdots h_{m_c}^\alpha+\cdots
    \nonumber\\
    &&+(-)^{c-1}B_c h_{m_1+\cdots+m_c}^\alpha\Biggr)
    \Biggr|_{h_m^\alpha=(z_\alpha)^m
    +\sum_{k=1}^{c-1}(\mu_k^\alpha)^m}
    \Biggr]\Biggr\}\Biggr\rangle,
\end{eqnarray}
and $\sD_P$ is a constant having the same expression
as that of $\sN_P$, except that
$k$ runs from 1 to $c$ and $z^\alpha$ are set to 0.

The expression in the exponential argument of $\sN_P$
can be further simplified.
Rewriting $\phi$ as unrestricted sums over the neighbours
analogously to Eq.~(\ref{eq:factorsum}),
\begin{eqnarray}
    \phi(q^\alpha,\mu_1^\alpha,\cdots,\mu_c^\alpha)
    &=&\frac{1}{c!}\sum_{\mathbf m}\phi_{\mathbf m}
    (q^\alpha)^{m_0}\left\{
    \left[\sum_{k=1}^c(\mu_k^\alpha)^{m_1}\right]\cdots
    \left[\sum_{k=1}^c(\mu_k^\alpha)^{m_c}\right]\right.
    \nonumber\\
    &&-B_2\left[\sum_{k=1}^c(\mu_k^\alpha)^{m_1+m_2}\right]
    \left[\sum_{k=1}^c(\mu_k^\alpha)^{m_3}\right]
    \cdots
    \left[\sum_{k=1}^c(\mu_k^\alpha)^{m_c}\right]+\cdots
    \nonumber\\
    &&\left.+(-)^{c-1}B_c
    \left[\sum_{k=1}^c(\mu_k^\alpha)^{m_1+\cdots+m_c}\right]
    \right\}.
\end{eqnarray}
Identifying each term in the square bracket as $h_1^\alpha,\cdots,h_Q^\alpha$,
we recognise the exponential argument as
$-\beta\sum_\alpha\phi(q^\alpha,z^\alpha,\mu_1^\alpha,
\cdots,\mu_{c-1}^\alpha)$.
We can now identify a recursion relation
for the function $R$
which does not involve replica indices,
\begin{equation}
    R(z,q|{\mathbf T})=\frac{1}{\sD_R}
    \prod_{k=1}^{c-1}\left[{\rm Tr}_{\bmu_k}
    R(q,\mu_k|{\mathbf T}_k)\right]
    \exp[-\beta\phi(q,z,\mu_1,\cdots,\mu_{c-1})].
\end{equation}
The denominator is given, in the limit $n$ approaching 0,
\begin{equation}
    \sD_R=\exp\left\langle\ln\left\{
    {\rm Tr}_{q,\bmu_k}
    \prod_{k=1}^c\left[R(q,\mu_k|{\mathbf T}_k)\right]
    \exp[-\beta\phi(q,\mu_1,\cdots,\mu_c)]\right\}\right\rangle.
\end{equation}
Letting the vertex free energy be defined by
$F^V(z,q|{\mathbf T})=-T\ln R(z,q|{\mathbf T})$,
we arrive at the recursion relation (\ref{eq:vertexfreerecursion_al})
and the average free energy (\ref{eq:fav}).

\section{Free Energy and Energy in the Paramagnetic State}

The average free energy is given by
\begin{equation}
\label{eq:Fav_noninteger}
    F_{\rm av}=P(C_j =3)\left.F_{\rm av}\right|_{C=3}
    +P(C_j =4)\left.F_{\rm av}\right|_{C=4}
    -\frac{\langle c\rangle}{2}\sum\limits_{C_iC_j}
    \frac{C_iP(C_i)}{\langle c\rangle}
    \frac{C_jP(C_j)}{\langle c\rangle}
    \left.F_{\rm link}\right|_{C_i C_j},
\end{equation}
where
\begin{eqnarray}
    \left. F_{\rm av} \right|_{C=3}
    &=& 4-\left\langle T\ln Q\left\{
    Q_1 Q_2 Q_3 + 3Q_1 Q_2z^2 + Q_1z^6 \right.\right.
    \nonumber \\
    &&+[Q_1Q_2z^2+Q_1z^4](z_{j1} +z_{j2}+z_{j3})
    \nonumber\\
    &&\left.\left.+Q_1z^6(z_{j1}z_{j2}+z_{j2}z_{j3}+z_{j1}z_{j3})
    +z^{12}z_{j1}z_{j2}z_{j3}\right\}\right\rangle \ ,
    \nonumber \\
    \left. {F_{\rm av} } \right|_{C=4} &=& 5-\left\langle T\ln
    Q\left\{ {Q_1Q_2Q_3Q_4}
    + 6Q_1Q_2Q_3z^2 +
    3Q_1Q_2z^4+4Q_1Q_2z^6+Q_1z^{12}\right.\right.
    \nonumber\\
    &&+[Q_1Q_2Q_3z^2+3Q_1Q_2z^4
    + Q_1z^8](z_{j1} +z_{j2} +z_{j3} +z_{j4})
    \nonumber \\
    &&+[Q_1Q_2z^6+Q_1z^8](z_{j1}z_{j2}+z_{j1}z_{j3}+z_{j1}z_{j4}
    +z_{j2}z_{j3}+z_{j2}z_{j4}+z_{j3}z_{j4})
    \nonumber \\
    &&+\left.\left. Q_1z^{12}(z_{j1} z_{j2} z_{j3} +z_{j1} z_{j2} z_{j4}
    +z_{j1} z_{j3} z_{j4} +z_{j2} z_{j3} z_{j4} )
    +z^{20}z_{j1} z_{j2} z_{j3} z_{j4}  \right\}
    \right\rangle ,
    \nonumber \\
     \left. {F_{\rm link} } \right|_{C_i C_j } &=& \left.
    {\left\langle {-T\ln Q\left[ {Q-1+z_{ij} z_{ji} } \right]}
    \right\rangle } \right|_{C_i C_j }  \ .
\label{eq:para4Fav}
\end{eqnarray}

The average energy is given by
\begin{equation}
\label{eq:paraEav_noninteger}
    E_{\rm av} =P(C_j =3)\left. {E_{\rm av} }
    \right|_{C=3} +P(C_j =4)\left. {E_{\rm av} } \right|_{C=4} ,
\end{equation}
the components of which take the form
\begin{equation}
    \left. {E^{(3)}_{\rm av} } \right|_{C=3} =\left\langle
    {\frac{E^{(3)}_N}{E^{(3)}_D}} \right\rangle  ,\mbox{~and~} \left.
    {E^{(4)}_{\rm av} } \right|_{C=4} =\left\langle
    {\frac{E^{(4)}_N}{E^{(4)}_D}} \right\rangle,
\end{equation}
where
\begin{eqnarray}
\label{eq:para3Eav}
    E^{(3)}_D&=&Q_1Q_2Q_3+3Q_1Q_2z^2+Q_1z^6
    +[Q_1Q_2z^2+Q_1z^4](z_{j1} +z_{j2} +z_{j3} )
    \nonumber \\
    &&+ Q_1z^6(z_{j1} z_{j2} +z_{j2} z_{j3} +z_{j1}
    z_{j3} )+z^{12}z_{j1} z_{j2} z_{j3} ,
    \nonumber\\
    E^{(3)}_N &=& 4Q_1Q_2Q_3+18Q_1Q_2z^2+10Q_1z^6
    + [6Q_1Q_2z^2+8Q_1z^4](z_{j1} +z_{j2} +z_{j3})
    \nonumber \\
    &&+ 10Q_1z^6(z_{j1} z_{j2} +z_{j2} z_{j3} +z_{j1}
    z_{j3} )+16z^{12}z_{j1} z_{j2} z_{j3}  \ ,
    \nonumber \\
    E^{(4)}_D&=& Q_1Q_2Q_3Q_4+6Q_1Q_2Q_3z^2
    +Q_1Q_2z^4+4Q_1Q_2z^6+Q_1z^{12}
    \nonumber \\
    &&+ [Q_1Q_2Q_3z^2+3Q_1Q_2z^4
    + Q_1z^8](z_{j1} +z_{j2} +z_{j3} +z_{j4} )
    \nonumber \\
    &&+ [Q_1Q_2z^6+ Q_1z^8](z_{j1} z_{j2} +z_{j1} z_{j3} +z_{j1}
    z_{j4} +z_{j2} z_{j3} +z_{j2} z_{j4} +z_{j3} z_{j4} )
    \nonumber \\
    &&+ Q_1z^{12}(z_{j1} z_{j2} z_{j3} +z_{j1} z_{j2} z_{j4} +z_{j1}
    z_{j3} z_{j4} +z_{j2} z_{j3} z_{j4})
    +z^{20}z_{j1} z_{j2} z_{j3} z_{j4} ,
    \nonumber \\
    E^{(4)}_N &=& 5Q_1Q_2Q_3Q_4+42Q_1Q_2Q_3z^2
    + 27Q_1Q_2z^4+44Q_1Q_2z^6+17Q_1z^{12}
    \nonumber \\
    &&+ [7Q_1Q_2Q_3z^2+27Q_1Q_2z^4
    + 13Q_1z^8](z_{j1} +z_{j2} +z_{j3} +z_{j4} )
    \nonumber\\
    &&+ [11Q_1Q_2z^6 + 13Q_1z^8](z_{j1} z_{j2} +z_{j1} z_{j3} +z_{j1}
    z_{j4} +z_{j2} z_{j3} +z_{j2} z_{j4} +z_{j3} z_{j4} )
    \nonumber \\
    &&+ 17Q_1z^{12}(z_{j1} z_{j2} z_{j3} +z_{j1} z_{j2} z_{j4}
    +z_{j1} z_{j3} z_{j4} +z_{j2} z_{j3} z_{j4} )
    + 25z^{20}z_{j1} z_{j2} z_{j3} z_{j4} \ .
    \nonumber
\end{eqnarray}


\section*{References}

\begin{thebibliography}{99}

\bibitem{sherrington1975} D. Sherrington and S. Kirkpatrick,
Phys. Rev. Lett. {\bf 35}, 1792 (1975).

\bibitem{kirkpatrick1978} S. Kirkpatrick and D. Sherrington,
Phys. Rev. B {\bf 17}, 4384 (1978).

\bibitem{nishimori2001}
H. Nishimori, {\em Statistical Physics of Spin Glasses and
Information Processing}, OUP UK (2001)

\bibitem{edwards1975} S. F. Edwards and P. W. Anderson,
J. Phys. F {\bf 5}, 965 (1975).

\bibitem{mezard1987}
M. M\'ezard, G. Parisi, and M. Virasoro, {\em Spin Glass Theory
and Beyond}, World Scientific, Singapore (1987)

\bibitem{fu1986} Y. Fu and P. W. Anderson,
J. Phys. A {\bf 19}, 1605 (1986).

\bibitem{mezard1986} M. M\'ezard and G. Parisi,
J. Physique {\bf 47}, 1285 (1986).

\bibitem{kirkpatrick1994} S. Kirkpatrick and B. Selman,
Science {\bf 264}, 1297 (1994).

\bibitem{mulet2002}
R. Mulet, A, Pagnani, M. Weigt, and R. Zecchina,
Phys. Rev. Lett. {\bf 89}, 268701 (2002).

\bibitem{COLbook}T. R. Jensen and B. Toft,
{\em Graph Coloring Problems} (Wiley Interscience, 1995).

\bibitem{NPbook} M. R. Garey and D. S. Johnson, {\em Computers and
Intractability} (Freeman, 1979).

\bibitem{braunstein2003}
A. Braunstein, R. Mulet, A. Pagnani, M. Weigt, and R. Zecchina,
Phys. Rev. E {\bf 68}, 036702 (2003).

\bibitem{vanmourik2002}
J. van Mourik and D. Saad,
Phys. Rev. E, {\bf 66}, 056120 (2002).

\bibitem{mezard2005} M. M\'ezard, M. Palassini, and O. Rivoire,
Phys. Rev. Lett. {\bf 95}, 200202 (2005).

\bibitem{monasson1999}
R. Monasson, R. Zecchina, S. Kirkpatrick, B. Selman, and L. Troyansky,
Nature {\bf 400}, 133 (1999).

\bibitem{krzakala2004}
F. Krzakala and A. Pagnani, Phys. Rev. E {\bf 70}, 046705 (2004).

\bibitem{krzakala2006}
F. Krzakala, A. Montanari, F. Ricci-Tersenghi, G. Semerjian,
and L. Zdeborov\'a, 
Proc. Natl. Acad. Sci. U. S. A. {\bf 104}, 10318 (2007).

\bibitem{zdeborova2007}
L. Zdeborov\'a and F. Krzakala, Phys. Rev. E {\bf 76}, 031131 (2007).

\bibitem{zdeborova2007a}
F. Krzakala and L. Zdeborov\'a, Europhys. lett. {\bf 81}, 57005 (2008).

\bibitem{frey1998} B. J. Frey,
Graphical Models for Machine Learning and Digital Communication
(MIT Press, Cambridge, MA, 1998).

\bibitem{mezard2002} M. M\'ezard, G. Parisi, and R. Zecchina,
Science {\bf 297}, 812 (2002).

\bibitem{mccormick1983}
S. T. McCormick, Math. Programming {\bf 26}, 153 (1983).

\bibitem{oceanstore} J. Kubiatowicz, D. Bindel, Y. Chen, P. Eaton, D. Geels,
R. Gummadi, S. Rhea, H. Weatherspoon, W. Weimer, C. Wells and B. Zhao,
OceanStore: An Extremely Wide-Area Storage System, 
U. C. Berkeley Technical Report No. UCB//CSD-00-1102 (1999).

\bibitem{bounkong2006}
S Bounkong, J van Mourik and D Saad, Phys. Rev. E {\bf 74}, 057101 (2006).

\bibitem{jiang2007}
A. Jiang and J. Bruck, Proc. IEEE Symposium on Information Theory, 381 (2002).

\bibitem{wong2006}
K. Y. M. Wong and D. Saad, Phys. Rev. E {\bf 74}, 010104 (2006).

\bibitem{wong2007}
K. Y. M. Wong and D. Saad, Phys. Rev. E {\bf 76}, 011115 (2007).

\bibitem{gross1985} D. J. Gross, I. Kanter, and H. Sompolinsky,
Phys. Rev. Lett. {\bf 55}, 304 (1985).

\bibitem{rivoire2004}
O. Rivoire, G. Biroli, O. C. Martin, and M. M\'ezard,
Eur. Phys. J. B {\bf 37}, 55 (2004).

\bibitem{mezard2001}
M. M\'ezard and G. Parisi, Eur. Phys. J. B {\bf 20}, 217 (2001).

\bibitem{monasson1997} R. Monasson and R. Zecchina,
Phys. Rev. E {\bf 56}, 1357 (1997).

\bibitem{wong1988}
K. Y. M. Wong, D. Sherrington, P. Mottishaw, R. Dewar,
and C. de Dominicis, J. Phys. A {\bf 21}, L99 (1988).

\end{thebibliography}
\end{document}